\documentclass[10pt,letterpaper]{article}
\usepackage[top=0.85in,left=1.00in,footskip=0.75in]{geometry}
\usepackage[utf8]{inputenc}
\usepackage{amsmath,amssymb}

\usepackage{mathtools}

\usepackage{changepage}

\usepackage{textcomp,marvosym}

\usepackage{cite}

\usepackage{nameref}
\usepackage[colorlinks=true, urlcolor=cuteBlue, citecolor=black, linkcolor=black]{hyperref}

\usepackage[right]{lineno}

\usepackage{microtype}
\DisableLigatures[f]{encoding = *, family = * }

\usepackage{array}

\usepackage{color}
\usepackage[table,xcdraw,dvipsnames]{xcolor}
\definecolor{cuteBlue}{rgb}{0.258, 0.387, 0.574}

\usepackage{xparse}

\DeclareDocumentCommand \eref{oooo} {\IfNoValueTF{#2}{Eq~(\ref{#1})}{\IfNoValueTF{#3}{Eqs~(\ref{#1}) and (\ref{#2})}{\IfNoValueTF{#4}{Eqs~(\ref{#1})-(\ref{#3})}{Eqs~(\ref{#1})-(\ref{#4})}}}}

\DeclareDocumentCommand \tref{oooo} {\IfNoValueTF{#2}{Table~\ref{#1}}{\IfNoValueTF{#3}{Tables~\ref{#1} and \ref{#2}}{\IfNoValueTF{#4}{Tables~\ref{#1}-\ref{#3}}{Tables~\ref{#1}-\ref{#4}}}}}

\DeclareDocumentCommand \fref{ooo} {\IfNoValueTF{#2}{Fig~\ref{#1}}{\IfNoValueTF{#3}{Figs~\ref{#1} and \ref{#2}}{Figs~\ref{#1}-\ref{#3}}}}

\usepackage{float}

\usepackage{hhline}

\newcommand{\letter}[1]{#1} 
\newcommand{\letterParen}[1]{(#1)} 

\usepackage{multirow}

\usepackage{units, xfrac}

\mathchardef\mhyphen="2D

\usepackage{accents}

\usepackage[aboveskip=1pt,labelfont=bf,labelsep=period,justification=raggedright,singlelinecheck=off]{caption}

\makeatletter
\renewcommand{\@biblabel}[1]{\quad#1.}
\makeatother

\begin{document}
\vspace*{0.2in}

\begin{flushleft}
{\Large
\textbf\newline{When two are better than one: Modeling the mechanisms of antibody mixtures}
}
\newline
\\
Tal Einav\textsuperscript{1},
Jesse D. Bloom\textsuperscript{1,2*}
\\
\bigskip
$^{1}\,$Basic Sciences Division and Computational Biology Program, Fred Hutchinson Cancer Research Center, Seattle, WA, United States of America
\\
$^{2}\,$Howard Hughes Medical Institute, Seattle, WA, United States of America
\\
\bigskip

* jbloom@fredhutch.org

\end{flushleft}


\section*{Abstract}

It is difficult to predict how antibodies will behave when mixed together, even
after each has been independently characterized. Here, we present a statistical
mechanical model for the activity of antibody mixtures that accounts for
whether pairs of antibodies bind to distinct or overlapping epitopes. This model
requires measuring $n$ individual antibodies and their $\frac{n (n-1)}{2}$
pairwise interactions to predict the $2^n$ potential combinations. We apply this
model to epidermal growth factor receptor (EGFR) antibodies and find that the
activity of antibody mixtures can be predicted without positing synergy at the molecular level. In
addition, we demonstrate how the model can be used in reverse, where
straightforward experiments measuring the activity of antibody mixtures can be
used to infer the molecular interactions between antibodies. Lastly, we
generalize this model to analyze engineered multidomain antibodies, where
components of different antibodies are tethered together to form novel amalgams,
and characterize how well it predicts recently designed influenza antibodies.


\section*{Author summary}

With the rise of new combination antibody therapeutic regimens, it is important to understand how antibodies work together as well as individually. Here, we investigate the specific case of monoclonal
antibodies targeting a cancer-causing receptor or the influenza virus and
develop a statistical mechanical framework that predicts the effectiveness of a
mixture of antibodies. The power of this model lies in its ability to
make a large number of predictions based on a limited amount of data. For
example, once 10 antibodies have been individually characterized, our model can predict how any of the $2^{10} = 1024$ combinations will behave. This
predictive power provides ample opportunities to test our model and paves the
way to expedite the design of future therapeutics.


\section*{Introduction}

Antibodies can bind with strong affinity and exquisite specificity to a
multitude of antigens. Due to their clinical and commercial success, antibodies
are one of the largest and fastest growing classes of therapeutic
drugs \cite{Awwad2018}. While most therapies currently use monoclonal
antibodies (mAbs), mounting evidence suggests
that mixtures of antibodies can behave in fundamentally different ways
\cite{Caskey2019}. There is ample precedent for the idea that combinations of therapeutics can be extremely powerful---for instance, during the past 50 years the monumental triumphs
of combination anti-retroviral therapy and chemotherapy cocktails have provided
unprecedented control over HIV and multiple types of cancer \cite{Perelson1997,
	Mukherjee2010}, and in many cases no single drug has emerged with comparable
effects. However, it is difficult to predict how antibody mixtures will
behave relative to their constitutive parts. Often, the vast number of potential
combinations is prohibitively large to systematically test, since both the
composition of the mixture and the relative concentration of each component can
influence its efficacy \cite{Chow2013}.

Here, we develop a statistical mechanical model that bridges the gap between how
an antibody operates on its own and how it behaves in concert. Specifically,
each antibody is characterized by its binding affinity and potency, while its
interaction with other antibodies is described by whether its epitope is distinct
from or overlaps with theirs. This information enables us to translate the
molecular details of how each antibody acts individually into the macroscopic
readout of a system's activity in the presence of an arbitrary mixture.

To test the predictive power of our framework, we apply it to a beautiful recent
case study of inhibitory antibodies against the epidermal growth factor receptor
(EGFR), where 10 antibodies were individually characterized for their ability to
inhibit receptor activity and then all possible 2-Ab and 3-Ab mixtures were
similarly tested \cite{Koefoed2011}. We demonstrate that our framework can
accurately predict the activity of these mixtures based solely on the behaviors of the ten
monoclonal antibody as well as their epitope mappings.

Lastly, we generalize our model to predict the potency of engineered multidomain
antibodies from their individual components. Specifically, we consider the
recent work by Laursen \textit{et al.} where four single-domain antibodies were
assayed for their ability to neutralize a panel of influenza strains, and then
the potency of constructs comprising 2-4 of these single-domain antibodies were
measured \cite{Laursen2018}. Our generalized model can once again
predict the efficacy of the multidomain constructs based upon their constitutive
components, once a single fit parameter is inferred to quantify the effects of
the linker joining the single-domain antibodies. This enables us to
quantitatively ascertain how tethering antibodies enhances the two key features
of potency and breadth that are instrumental for designing novel anti-viral
therapeutics. 
Notably, our models do \emph{not} posit complex molecular synergy between antibodies.
Our results therefore show that many antibody mixtures function without synergy, and hence that their effects can be
computationally predicted to expedite future experiments.

\section*{Results}

\subsection*{Modeling the mechanisms of action for antibody mixtures}

Consider a monoclonal antibody that binds to a receptor and inhibits its
activity. Two parameters characterize this inhibition: (1) the dissociation
constant $K_D$ quantifies an antibody's binding affinity (with a smaller value
indicating tighter binding) and (2) the potency $\alpha$ relates the activity
when an antibody is bound to the activity in the absence of antibody. A value of $\alpha =
1$ represents an impotent antibody that does not affect activity while $\alpha =
0$ implies that an antibody fully inhibits activity upon binding. As derived in
S1 Text Section A, for an antibody that binds to a single site
on a receptor, the activity at a concentration $c$ of antibody is given by
\begin{equation} \label{eqActivity1Ab}
\text{Fractional Activity} = \frac{1 + \alpha \frac{c}{K_D}}{1 + \frac{c}{K_D}}.
\end{equation}

To characterize a mixture of two antibodies, we not only need their individual
dissociation constants and potencies but also require a model for how these
antibodies interact. When two antibodies bind to distinct epitopes, the simplest
scenario is that their ability to bind and inhibit activity is independent of
the presence of the other antibody, and hence that their combined potency when
simultaneously bound equals the product of their individual potencies
(\fref[figAntibodyMixtureModels]\letter{A}). Alternatively, if the two
antibodies compete for the same epitope, they cannot both be simultaneously
bound (\fref[figAntibodyMixtureModels]\letter{B}).


We also define the general case of a synergistic interaction where the binding
of the first antibody alters the binding or potency of the
second antibody (\fref[figAntibodyMixtureModels]\letter{C}, purple text). This
definition encompasses cases where the second antibody binds more tightly
($K_{D,\text{eff}}^{(2)} < K_{D}^{(2)}$) or more weakly ($K_{D,\text{eff}}^{(2)}
> K_{D}^{(2)}$) in the presence of the first antibody, as well as when the
potency of the second antibody may increase ($\alpha_{2,\text{eff}} > \alpha_2$)
or decrease ($\alpha_{2,\text{eff}} < \alpha_2$). This also includes cases where two
epitopes slightly overlap and partially inhibit one another's binding, and the
competitive binding model can be viewed as the extreme limit
$K_{D,\text{eff}}^{(2)} \to \infty$ where one antibody infinitely penalizes the
binding of the other.

\begin{figure}[t!]
	\centering \includegraphics{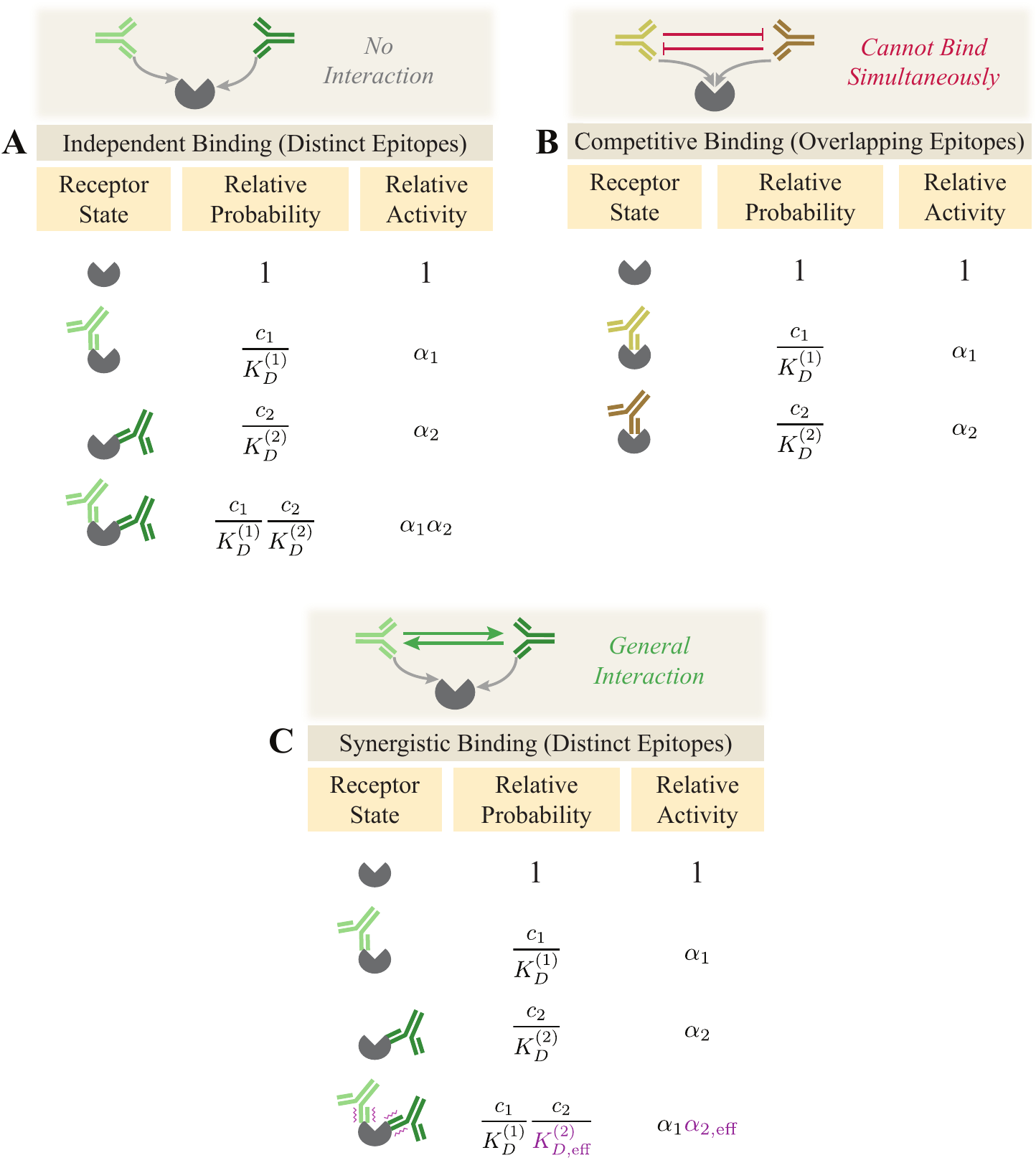}
	
	\caption{\textbf{Binding modes for a 2-Ab mixture.} Two antibodies with
		concentrations $c_1$ and $c_2$ can bind \letterParen{A} independently to
		different epitopes or \letterParen{B} competitively to the same epitope.
		\letterParen{C} Antibodies bind synergistically if either the product of
		binding affinities ($K_D^{(j)}$) or potencies ($\alpha_j$) are altered
		when both antibodies bind.} \label{figAntibodyMixtureModels}
\end{figure}

While the synergistic model in \fref[figAntibodyMixtureModels]\letter{C} has the merit of being highly
general, an important feature of the independent and competitive models
(\fref[figAntibodyMixtureModels]\letter{A,B}) is that they predict all antibody
combinations with few parameters. In both of these latter models, once the
$K_D^{(j)}$ and $\alpha_j$ of 10 antibodies are known (which requires $2 \cdot
10$ experiments) and their epitopes are mapped ($\frac{10 \cdot 9}{2}$
additional experiments), the potency of all $2^{10}=1024$ possible mixtures of
these antibodies can be predicted without recourse to fitting. 
In contrast, because the synergistic model allows arbitrary interactions between each combination
of antibodies, the behavior of a mixture exhibiting synergy cannot be predicted without actually
making a measurement on that combination to quantify the synergy.

For these reasons, in this work we focus on the two cases of independent or
competitive binding and show how we can combine both models to transform our
molecular understanding of each monoclonal antibody's action into a prediction
of the efficacy of an antibody mixture. Deviations from our predictions provide
a rigorous way to measure antibody synergy by computing
$\frac{K_{D,\text{eff}}^{(2)}}{K_{D}^{(2)}}$ and
$\frac{\alpha_{2,\text{eff}}}{\alpha_2}$.

To mathematize the independent and competitive binding models, we enumerate the
possible binding states and compute their relative Boltzmann weights.
The fractional activity of each state equals the product of its
relative probability and relative activity divided by the sum of all relative
probabilities for normalization (see S1 Text Section
A). When two antibodies bind independently
as in \fref[figAntibodyMixtureModels]\letter{A}, this factors into the form
\begin{equation} \label{eqActivity2AbDistinct}
\text{Fractional Activity}_{\text{(distinct epitopes)}} = \left( \frac{1 + \alpha_1 \frac{c_1}{K_D^{(1)}}}{1 + \frac{c_1}{K_D^{(1)}}} \right) \left( \frac{1 + \alpha_2 \frac{c_2}{K_D^{(2)}}}{1 + \frac{c_2}{K_D^{(2)}}} \right).
\end{equation}
If these two antibodies compete for the same epitope as in \fref[figAntibodyMixtureModels]\letter{B}, the activity becomes
\begin{equation} \label{eqActivity2AbOverlapping}
\text{Fractional Activity}_{\text{(overlapping epitopes)}} = \frac{1 + \alpha_1 \frac{c_1}{K_D^{(1)}} + \alpha_2 \frac{c_2}{K_D^{(2)}}}{1 + \frac{c_1}{K_D^{(1)}} + \frac{c_2}{K_D^{(2)}}}.
\end{equation}
These equations are readily extended to mixtures with three or more antibodies (see S1 Text Section A).

\subsection*{Antibody mixtures against EGFR are well characterized using independent and competitive binding models}

To test the predictive power of the independent and competitive binding models,
we applied them to published experiments on the epidermal growth factor receptor
(EGFR) where ten monoclonal antibodies were individually characterized and then
the activity of all 165 possible 2-Ab and 3-Ab mixtures was measured
\cite{Koefoed2011}. We first use each monoclonal antibody's response to infer
its dissociation constant $K_D$ and potency $\alpha$. We then utilize surface
plasmon resonance (SPR) measurements to determine which pairs of antibodies bind
independently and which compete for the same epitope. These data enable us to
use the above framework and predict EGFR activity in the presence of any
mixture.


EGFR is a transmembrane protein that activates in the presence of epidermal
growth factors. Upon ligand binding, the receptor's intracellular tyrosine
kinase domain autophosphorylates which leads to downstream signaling cascades
central to cell migration and proliferation. Overexpression of EGFR has been
linked to a number of cancers, and decreasing EGFR activity in such tumors by
sterically occluding ligand binding has reduced the rate of cancer proliferation
\cite{Koefoed2011}.

Koefoed \textit{et al.}~investigated how a panel of ten monoclonal antibodies
inhibit EGFR activity in the human cell line A431NS \cite{Koefoed2011}. They
then measured how 1:1 mixtures of two antibodies or 1:1:1 mixtures of three
antibodies affect EGFR activity. All measurement were carried out at a total
concentration of $2 \frac{\mu \text{g}}{\text{mL}}$, implying that each antibody
was half as dilute in the 2-Ab mixtures and one-third as dilute in the 3-Ab
mixtures relative to the monoclonal antibody measurement.


The 45 possible 2-Ab mixtures (35 binding to distinct epitopes; 10 binding to
overlapping epitopes) and the 120 possible 3-Ab mixtures (50 binding to distinct
epitopes; 70 binding to overlapping epitopes) were assayed for their ability to
inhibit EGFR activity. \fref[figIndependentModel]\letter{A} shows the
experimental measurements for mixtures of two antibodies, with the monoclonal antibody
measurements shown on the diagonal, the measured activity of 2-Ab mixtures shown
on the bottom-left and the predicted activity on the top-right. Each antibody
is labeled with its binding epitopes inferred through SPR \cite{Koefoed2011}, so
that antibodies binding to overlapping epitopes are predicted using
\eref[eqActivity2AbOverlapping] (pairs within the dashed gray boxes) while
mixtures binding to distinct epitopes use \eref[eqActivity2AbDistinct].

For example, antibodies \#1 and \#4 bind to distinct epitopes (III/C and III/B,
respectively). Hence, the predicted activity of their mixture (0.50) very nearly
equals the product of their individual activity ($0.65 \times 0.75 = 0.49$),
with the slight deviation arising because each antibody concentration was halved
in the mixture ($c_1 = c_2 = 1 \frac{\mu \text{g}}{\text{mL}}$ for the 2-Ab
mixture characterized by \eref[eqActivity2AbDistinct], whereas the individual
mAbs were measured at $c = 2 \frac{\mu \text{g}}{\text{mL}}$ using
\eref[eqActivity1Ab]). The predicted activity roughly approximates the measured
value 0.43 of the mixture.

\begin{figure}[t!]
	\centering \includegraphics{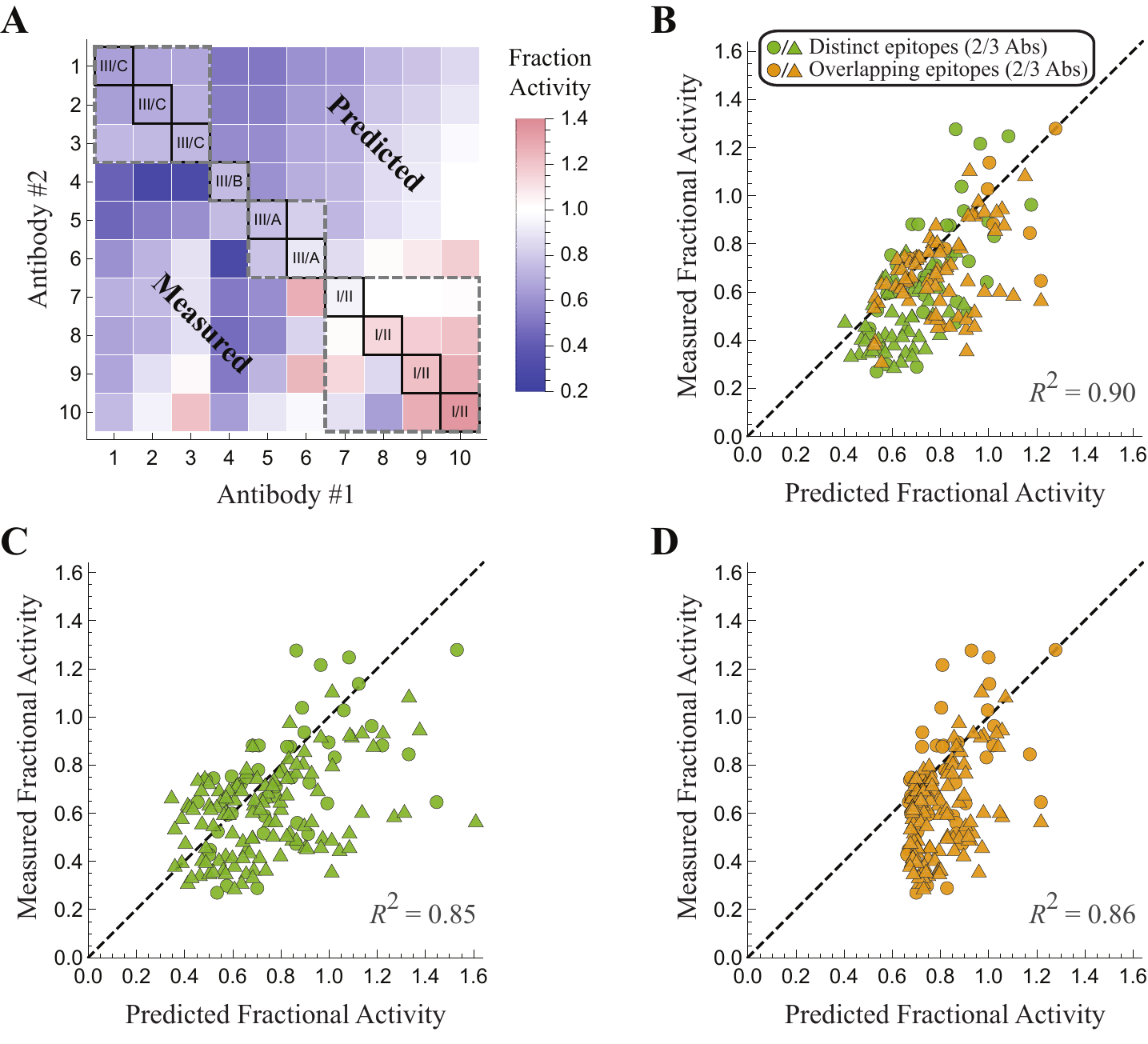}
	
	\caption{\textbf{Predicting how antibody mixtures affect the epidermal growth
		factor receptor (EGFR).} \letterParen{A} The fractional activity of EGFR in the
	presence of monoclonal antibodies (diagonal) together with the measured
	(bottom-left) and predicted (top-right) activity of all 2-Ab combinations. The
	dashed gray boxes enclose antibody pairs that compete for the same epitope
	while all other pairs bind independently. \letterParen{B} The predicted versus
	measured fractional activity for all 2-Ab and 3-Ab mixtures using the same
	epitope mapping as in Panel A inferred by SPR. Without the epitope map, the
	activity of the mixtures could alternately be predicted by assuming that all
	antibodies either \letterParen{C} bind independently or \letterParen{D} compete
	for the same epitope; in either case, the resulting predictions fall further
	from the diagonal line, indicating poorer predictive power.}
\label{figIndependentModel}
\end{figure}

On the other hand, antibodies \#1 and \#2 bind to the same epitope (III/C), and
hence their predicted combined activity (0.67) lies between their individual
activities (0.65 and 0.69) since both antibodies compete for the same site. The
measured activity of the mixture (0.65) closely matches the prediction of the overlapping epitope
model, but is very different than the prediction of 0.45 made by the distinct-binding model. 

\fref[figIndependentModel]\letter{B} shows the measured EGFR activity in the
presence of all 2-Ab and 3-Ab mixtures is highly correlated with the predicted activity ($R^2 = 0.90$)
Notably, the predictions are made solely from the monoclonal antibody data and epitope measurements, and do not involve any fitting of the 2-Ab or 3-Ab measurements.
The strong correlation between the predicted and measured activities suggests
that EGFR antibody mixtures can be characterized with minimal synergistic
effects in either their binding or effector functions. If we did not have the
epitope mapping through SPR and assumed that all antibodies bound to distinct
epitopes (\fref[figIndependentModel]\letter{C}, $R^2 = 0.85$) or competed for
the same epitope (\fref[figIndependentModel]\letter{D}, $R^2 = 0.86$), the
resulting predictions are slightly more scattered from the diagonal,
demonstrating that properly acknowledging which pairs of antibodies vie for the
same epitope boosts the predictive power of the model.

That said, the predictions incorporating the SPR mapping display a consistent
bias towards having a slightly lower measured than predicted activity, suggesting that several pairs of antibody enhance one another's
binding affinity or potency. To quantify this, if we recharacterize the activity
from the 2-Ab mixtures to a synergistic model where each $\alpha_{2,\text{eff}}$
is fit to exactly match the data, we find an average value of
$\frac{\alpha_{2,\text{eff}}}{\alpha_2} = 0.9$, showing that when pairs of
antibodies are simultaneously bound they typically boost their collective inhibitory activity by $\sim$10\%.

\subsection*{Differentiating distinct versus overlapping epitopes using antibody mixtures}

In the previous section, we used SPR measurements to quantify which antibodies
compete for overlapping epitopes, thereby permitting us to translate the
molecular knowledge of antibody interactions into a macroscopic quantity of
interest, namely, the activity of EGFR. In this section, we do the reverse and
utilize activity measurements to categorize which subsets of antibodies bind to
overlapping epitopes. This method can be applied to model antibody mixtures in
other biological systems where SPR measurements are not readily available.

For the remainder of this section, we ignore the known epitope mappings
discerned by Koefoed \textit{et al.}~and consider what mapping best
characterizes the data. For example, given the individual activities of antibody
\#1 (0.65) and \#2 (0.69), the predicted activity of their combination (at the
concentration of $1 \frac{\mu \text{g}}{\text{mL}}$ for each antibody dictated
by the experiments) would be 0.45 if they bind to distinct epitopes and 0.67 if
they bind to overlapping epitopes. Since the measured activity of this mixture
was 0.65, it suggests the latter option. We note that such analysis will work
best for potent antibodies (whose individual activity is far from 1), since only in this regime will the predictions of the distinct versus
overlapping models be significantly different. Therefore, the activity
measurements of each individual antibody would optimally be carried out at saturating
concentrations (where \eref[eqActivity1Ab] is as far from 1 as possible).

Proceeding to the other antibodies, we characterize each pair according
to whichever model prediction lies closer to the experimental measurement. To
account for experimental error, we left an antibody pair uncategorized if the
two model predictions were too close to one another (within $4\sigma = 0.16$
where $\sigma$ is the SEM of the measurements) or if the experimental
measurement was close (within $1\sigma$) to the average of the two model
predictions (see S1 Text Section B).

\fref[figPredictingDistinctVersusOverlappingEpitopes]\letter{A} shows how this
analysis compares to the experimental measurement inferred by SPR. While the
model predictions are much sparser (with the majority of antibody pairs
uncategorized because the two model predictions were too close to one another),
the classifications only disagreed with the SPR measurements in two cases
(claiming that antibodies \#7-8 overlap with antibody \#10; notice that
antibodies \#7-8 have individual activities close to 1, making them difficult to
characterize). 

Using these classifications, we defined unique EGFR epitopes by grouping
together any antibodies that bind to overlapping epitopes. In this way, we split
the ten antibodies into four distinct groups (antibodies \#1-3, \#4-5, \#6, and
\#7-10 indicated by the dashed gray rectangles in
\fref[figPredictingDistinctVersusOverlappingEpitopes]\letter{A}),
enabling us to distinguish which antibodies bind independently or competitively
and hence predict the activity of the 2-Ab and 3-Ab mixtures. Note that it is not the
pairwise classification between two antibodies that determines whether we apply
the distinct or competitive models, but rather these four groupings of antibody
epitopes. For example, although antibodies \#7 and \#8 are uncategorized through
their 2-Ab mixture, they fall within a single epitope group and hence are
considered to bind competitively. Similarly, antibody \#1 and \#4 are modeled as
binding independently because they belong to two distinct epitope groups.
Antibody \#6 is considered to be in its own epitope group since it did not
overlap with any other antibody.


Surprisingly, the results shown in
\fref[figPredictingDistinctVersusOverlappingEpitopes]\letter{B} have a
coefficient of determination $R^2=0.90$ that is on par with the results obtained
using the SPR measurements (\fref[figIndependentModel]\letter{B}). 
This suggests that there is no
loss in the predictive power of the model when an epitope mapping is inferred
through activity measurements.

\begin{figure}[t!]
	\centering \includegraphics{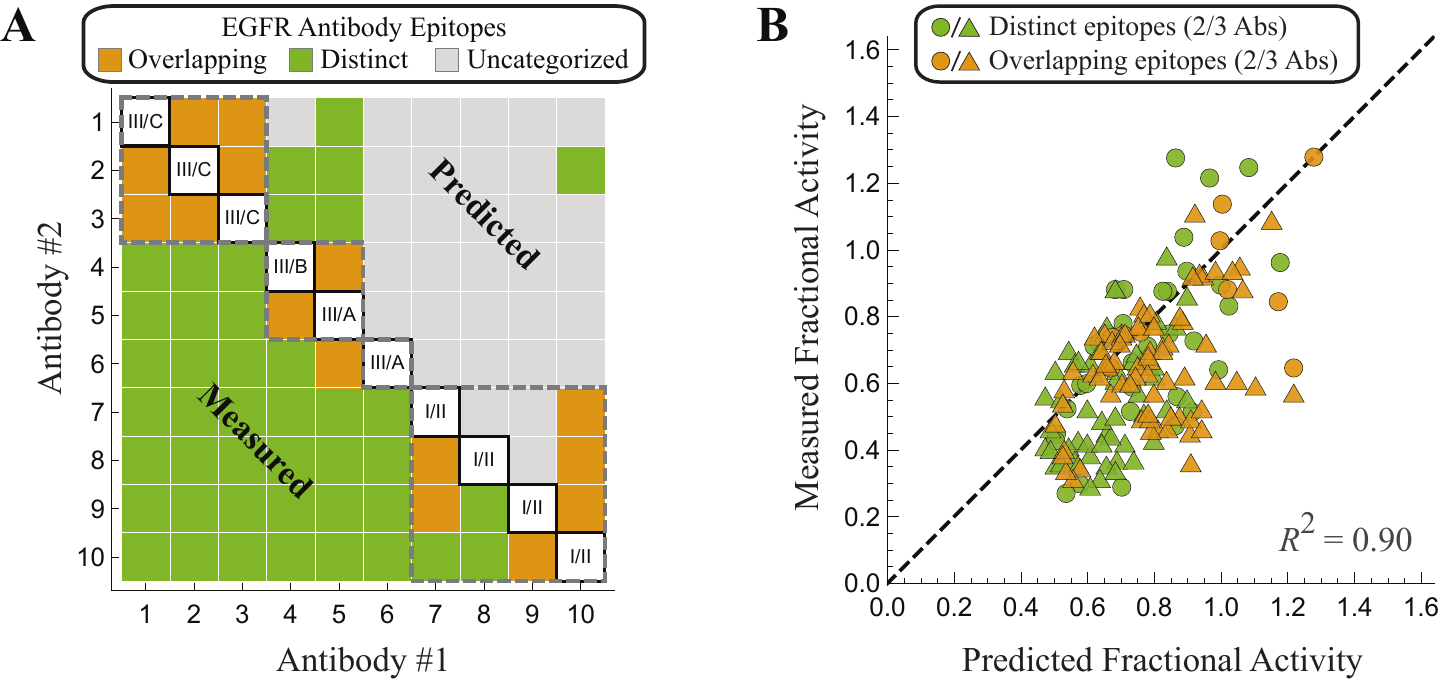}
	
	\caption{\textbf{Classifying antibody epitopes as overlapping or distinct.}
		\letterParen{A} Comparing the experimentally measured activity to the
		overlapping or distinct epitope models enables us to characterize each antibody
		pair (provided the two models predict sufficiently different activities).
		\letterParen{B} The resulting predictions for the 2-Ab and 3-Ab mixtures have
		the same predictive power ($R^2 = 0.90$) as a model that relies on epitope
		groupings given by SPR measurements (\fref[figIndependentModel]\letter{B}).}
	\label{figPredictingDistinctVersusOverlappingEpitopes}
\end{figure}

In summary, whether antibodies bind independently or competitively can be
determined either: (1) directly through pairwise competition experiments or (2)
by analyzing the activity of their 2-Ab mixtures in light of our two models.
When this information is combined with the potency and dissociation constant of
each antibody, the activity of an arbitrary mixture can be predicted. The
Supplementary Information contains a Mathematica program that can analyze either
form of the pairwise interactions to determine the epitope grouping. If the
characteristics of the individual antibodies are also provided, the program can
predict the activity of any antibody mixtures at any specified ratio of the
constituents.

\subsection*{Multidomain antibodies boost breadth and potency via avidity}

While the previous sections analyzed combinations of whole, unmodified
antibodies, we now extend our framework to connect with the rising tide of
engineering efforts that genetically fuse different antibody components to
construct multi-domain antibodies \cite{Spiess2015}. Specifically, we focus our
attention on recent work by Laursen \textit{et al.}~who isolated single-domain
antibodies from llamas immunized with H2 or H7 influenza hemagglutinin (HA)
\cite{Laursen2018}. The four single-domain antibodies isolated in this manner
included one antibody that preferentially binds influenza A group 1 strains
($\text{Ab}_{\text{A1}}$), another that binds influenza A group 2 strains
($\text{Ab}_{\text{A2}}$), and two antibodies that bind to influenza B strains
($\text{Ab}_{\text{B}}^{(1)}$ and $\text{Ab}_{\text{B}}^{(2)}$).
\fref[figTetheredAntibodyModel]\letter{A,B} shows data from a representative
influenza A group 1 strain (blue dot, only bound by the blue
$\text{Ab}_{\text{A1}}$), influenza A group 2 strain (green dot, only bound by
the green $\text{Ab}_{\text{A2}}$), and influenza B strain (gold dot, bound by
both of the yellow $\text{Ab}_{\text{B}}^{(1)}$ and $\text{Ab}_{\text{B}}^{(2)}$
antibodies). 

In the contexts of rapidly evolving pathogens such as influenza, two important
characteristics of antibodies are their potency and breadth. Potency is measured
by the inhibitory concentration $\text{IC}_{50}$ at which 50\% of a virus is
neutralized, where a smaller $\text{IC}_{50}$ represents a better antibody.
Breadth is a measure of how many strains are susceptible to an antibody.

In an effort to improve the potency and breadth of their antibodies, Laursen
\textit{et al.}~tethered together different domains using a flexible amino acid
linker (right-most columns of \fref[figTetheredAntibodyModel]\letter{A,B}) and
tested them against a panel of influenza strains. To make contact with these
multidomain constructs, consider a concentration $c$ of the tethered antibody
$\text{Ab}_{\text{A1}} \text{\textendash} \text{Ab}_{\text{A2}}$. Relative to
the unbound HA state, the $\text{Ab}_{\text{A1}}$ or $\text{Ab}_{\text{A2}}$
portions of the antibody will neutralize the virus with relative probability
$\frac{c}{\text{IC}_{50,\text{A1}}}$ or $\frac{c}{\text{IC}_{50,\text{A2}}}$,
respectively. Although neutralization is mediated by antibody binding, the two
quantities may or may not be proportional \cite{Knossow2002, Ndifon2009,
	Einav2019}, and hence we replace dissociation constants with $\text{IC}_{50}$s
in our model (see S1 Text Section C).

Laursen \textit{et al.}~determined that their tethered constructs cannot
intra-spike crosslink two binding sites on a single HA trimer, but they can
inter-spike crosslink adjacent HA \cite{Laursen2018}. The linker connecting the
two antibody domains facilitates such crosslinking, since when one domain is
bound the other domain is confined to a smaller volume around its potential
binding sites. This effect can be quantified by stating that the second domain
has an effective concentration $c_{\text{eff}}$
(\fref[figTetheredAntibodyModel]\letter{C}, purple), making the relative
probability of the doubly bound state $\frac{c}{\text{IC}_{50,\text{A1}}}
\frac{c_{\text{eff}}}{\text{IC}_{50,\text{A2}}}$. Therefore, the fraction of
virus neutralized by two tethered antibody domains is given by
\begin{equation} \label{eqFractionNeutralized}
\text{Fraction Neutralized} = \frac{\frac{c}{\text{IC}_{50,\text{A1}}} + \frac{c}{\text{IC}_{50,\text{A2}}} +
	\frac{c}{\text{IC}_{50,\text{A1}}} \frac{c_{\text{eff}}}{\text{IC}_{50,\text{A2}}}}{1 +
	\frac{c}{\text{IC}_{50,\text{A1}}} + \frac{c}{\text{IC}_{50,\text{A2}}} +
	\frac{c}{\text{IC}_{50,\text{A1}}} \frac{c_{\text{eff}}}{\text{IC}_{50,\text{A2}}}}.
\end{equation}
Note that this equation assumes that influenza virus is fully neutralized at
saturating concentrations of antibody ($\alpha = 0$ in \eref[eqActivity1Ab],
with $\text{Fraction Neutralized}$ analogous to $1 - \text{Fractional
	Activity}$).


\begin{figure}[t!]
	\centering \includegraphics{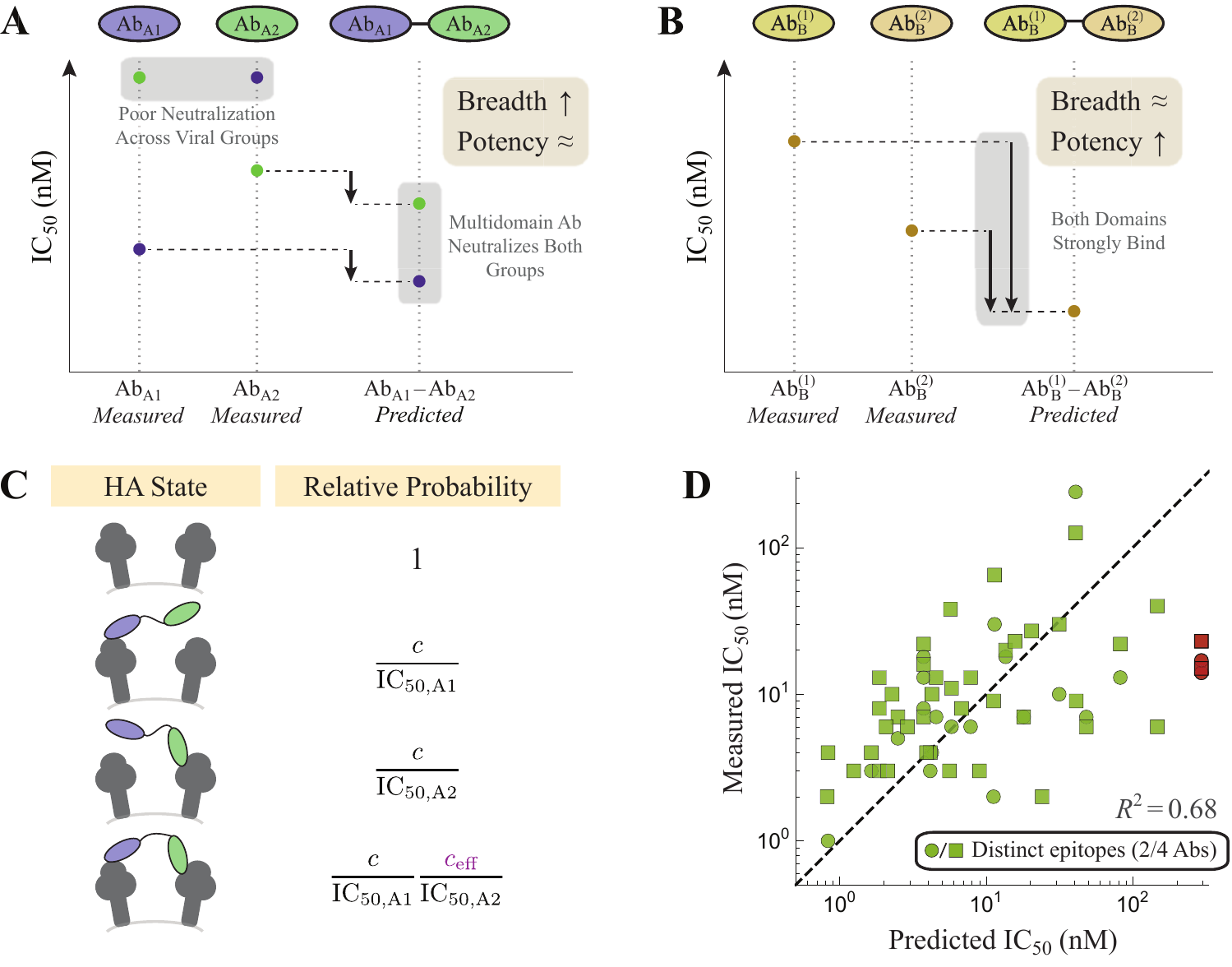}
	
	\caption{\textbf{Tethering influenza antibodies increases breadth and potency.}
		\letterParen{A} The influenza A antibodies $\text{Ab}_{\text{A1}}$ and
		$\text{Ab}_{\text{A2}}$ were tethered together to form $\text{Ab}_{\text{A1}}
		\text{\textendash} \text{Ab}_{\text{A2}}$ while \letterParen{B} two influenza B
		antibodies formed $\text{Ab}_{\text{B}}^{(1)} \text{\textendash}
		\text{Ab}_{\text{B}}^{(2)}$. Representative data shown for an influenza A group
		1 (blue), influenza A group 2 (green), and influenza B (gold) strains. Strong
		potency is marked by a small $\text{IC}_{50}$ while large breadth implies that
		multiple strains are controlled by an antibody. \letterParen{C} Representative states of HA and their
		corresponding Boltzmann weights for multidomain antibodies, where crosslinking
		between adjacent spikes boosts neutralization via avidity ($c_{\text{eff}} =
		1400\,\text{nM}$ in \eref[eqIC50TwoFusedAntibodies]). \letterParen{D}
		Theoretical predictions of the potency of all multidomain antibodies versus
		their measured values. The red points denote two outlier
		influenza strains discussed in the text that are not neutralized by
		$\text{Ab}_{\text{A1}}$ or $\text{Ab}_{\text{A2}}$ individually but are highly
		neutralized by their combination.} \label{figTetheredAntibodyModel}
\end{figure}

The $\text{IC}_{50}$ of the tethered construct is defined as the concentration $c$
at which half of the virus is neutralized, which can be solved to yield
\begin{equation} \label{eqIC50TwoFusedAntibodies}
\text{IC}_{\text{50,A1\text{\textendash}A2}} = \frac{\text{IC}_{\text{50,A1}} \, \text{IC}_{\text{50,A2}}}{c_{\text{eff}} + \text{IC}_{\text{50,A1}} + \text{IC}_{\text{50,A2}}},
\end{equation} 
with an analogous expression holding for the $\text{Ab}_{\text{B}}^{(1)}
\text{\textendash} \text{Ab}_{\text{B}}^{(2)}$ construct. Using the measured
$\text{IC}_{50}$s of $\text{Ab}_{\text{A1}} \text{\textendash}
\text{Ab}_{\text{A2}}$ and $\text{Ab}_{\text{B}}^{(1)} \text{\textendash}
\text{Ab}_{\text{B}}^{(2)}$ against the various influenza strains, we can infer
the value of the single parameter $c_{\text{eff}} = 1400\,\text{nM}$ (see
S1 Text Section C). This result is both
physically meaningful and biologically actionable, as it enables us to predict
the $\text{IC}_{50}$ of the tethered multidomain antibodies against the entire
panel of influenza strains. \fref[figTetheredAntibodyMeasurements]\letter{A,B}
compares the resulting predictions to the experimental measurements, where plot
markers linked by horizontal line segments indicate a close match between the
predicted and measured values. 

The two tethered antibodies display unique trends that arise from their
compositions. Since the two domains in $\text{Ab}_{\text{A1}} \text{\textendash}
\text{Ab}_{\text{A2}}$ bind nearly complementary strains, the tethered construct
will increase breadth (since this multidomain antibodies can now bind to both
group 1 and group 2 strains) but will only marginally improve potency.
Mathematically, if $\text{Ab}_{\text{A1}}$ binds tightly to an influenza A group
1 strain while $\text{Ab}_{\text{A2}}$ binds weakly to this same strain
($\text{IC}_{\text{50,A2}} \to \infty$), their tethered construct has an
$\text{IC}_{\text{50,A1\text{\textendash}A2}} \approx \text{IC}_{\text{50,A1}}$.
Said another way, $\text{Ab}_{\text{A1}} \text{\textendash}
\text{Ab}_{\text{A2}}$ should be approximately as potent as a mixture of the
individual antibodies $\text{Ab}_{\text{A1}}$ and $\text{Ab}_{\text{A2}}$. Note
that since the experiments could not accurately measure weak binding
($>1000\,\text{nM}$), the predicted $\text{IC}_{50}$ for the multidomain
antibodies represents a lower bound.

On the other hand, tethering the two influenza B antibodies yields a marked
improvement in potency over either individual antibody, since both domains can
bind to any influenza B strain and boost neutralization via avidity. The process
of engineering a multivalent interaction is reminiscent of engineered bispecific
IgG \cite{Spiess2015}, and adding additional domains could yield further
enhancement in potency, provided that all domains can simultaneously bind.

\begin{figure}[t!]
	\centering \includegraphics{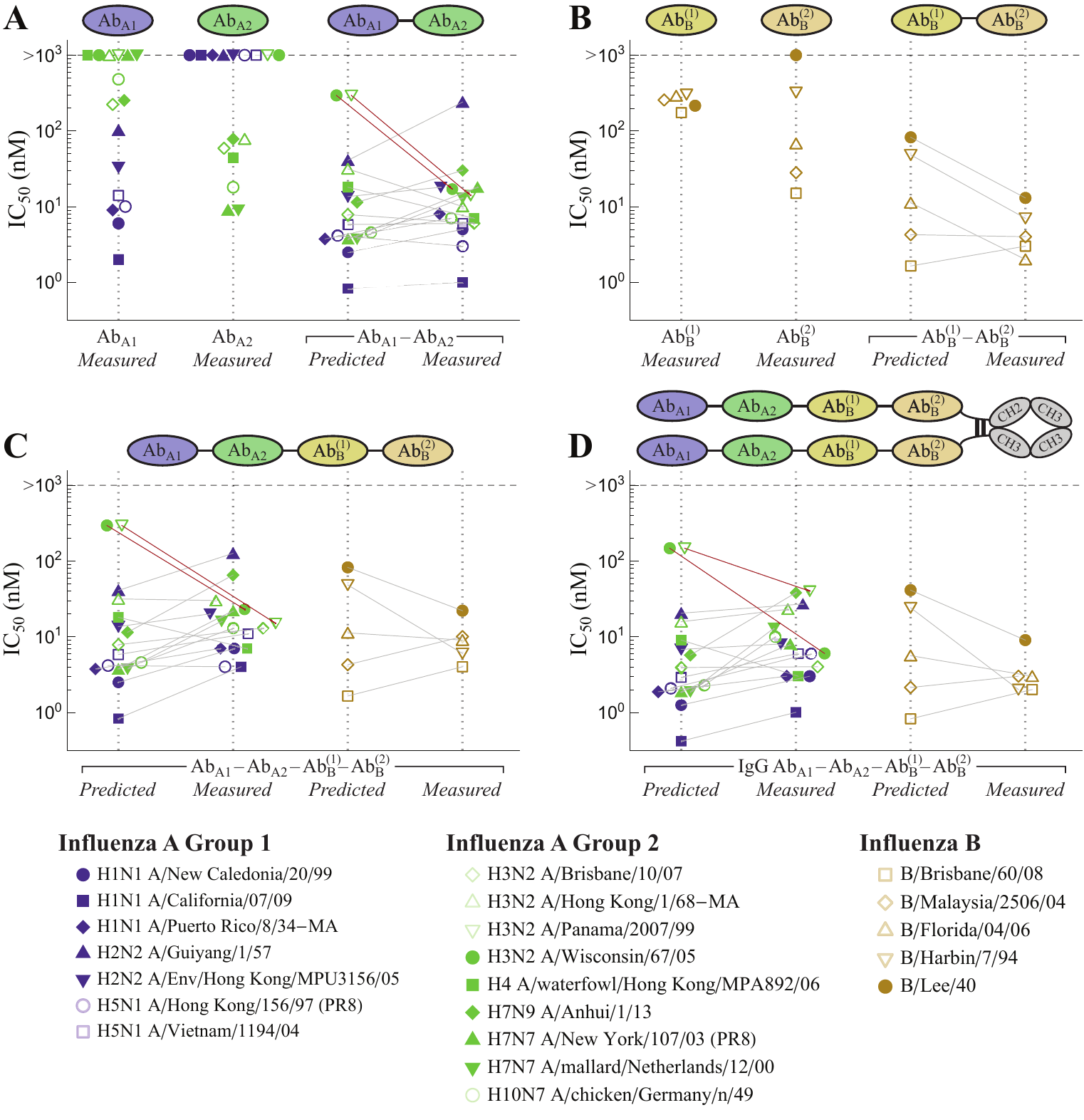}
	
	\caption{\textbf{Neutralization of multidomain antibodies.} \letterParen{A,B}
	The potency of the 2-Ab constructs and their constitutive antibodies against a
	panel of influenza strains. $\text{Ab}_{\text{A1}}$ primarily binds influenza A
	group 1 (blue), $\text{Ab}_{\text{A2}}$ to influenza A group 2 (green), and the
	two $\text{Ab}_{\text{B}}$ antibodies to influenza B strains (gold).
	\letterParen{C} All four antibodies were tethered to form the linear chain
	$\text{Ab}_{\text{A1}} \text{\textendash} \text{Ab}_{\text{A2}}
	\text{\textendash} \text{Ab}_{\text{B}}^{(1)} \text{\textendash}
	\text{Ab}_{\text{B}}^{(2)}$ and \letterParen{D} two copies of this chain were
	placed on an IgG backbone. The model suggests that the two arms of the IgG are
	not capable of simultaneously binding a virion. Red lines indicate two outlier
	influenza strains discussed in the text that are not neutralized by
	$\text{Ab}_{\text{A1}}$ or $\text{Ab}_{\text{A2}}$ individually but are highly
	neutralized by their combination. Data was digitized from Figs 1 and 3 of
	Ref~\cite{Laursen2018}.} \label{figTetheredAntibodyMeasurements}
\end{figure}

While the model is able to characterize the majority of tethered antibodies, it
also highlights some of the outliers in the data. For example, the H3N2 strains
A/Panama/2007/99 and A/Wisconsin/67/05 were poorly neutralized by either
$\text{Ab}_{\text{A1}}$ or $\text{Ab}_{\text{A2}}$ ($\text{IC}_{50} \ge
1000\,\text{nM}$), but the tethered construct exhibited an $\text{IC}_{50} =
14\,\text{nM}$ and $\text{IC}_{50} = 17\,\text{nM}$, respectively, far more
potent than the $300\,\text{nM}$ lower limit predicted for both viruses (red
circles in \fref[figTetheredAntibodyModel]\letter{D} and red lines in
\fref[figTetheredAntibodyMeasurements]\letter{A}). Interestingly, Laursen
\textit{et al.}~found that mixing the individual, untethered antibodies
$\text{Ab}_{\text{A1}}$ and $\text{Ab}_{\text{A2}}$ also resulted in shockingly
poor neutralization ($\text{IC}_{50} \ge 1000\,\text{nM}$), suggesting that the
tether is responsible for the increase in potency \cite{Laursen2018}. From the
vantage of our quantitative model, this outlier cries out for further
investigation.

To further boost neutralization, Laursen \textit{et al.}~created two additional
constructs that combined all four antibody domains, the first being the linear
chain ($\text{Ab}_{\text{A1}} \text{\textendash} \text{Ab}_{\text{A2}}
\text{\textendash} \text{Ab}_{\text{B}}^{(1)} \text{\textendash}
\text{Ab}_{\text{B}}^{(2)}$). Since the influenza A antibodies do not bind the
influenza B strains (and vise versa), this construct should have the same
$\text{IC}_{50}$ as $\text{Ab}_{\text{A1}} \text{\textendash}
\text{Ab}_{\text{A2}}$ for the influenza A strains and as
$\text{Ab}_{\text{B}}^{(1)} \text{\textendash} \text{Ab}_{\text{B}}^{(2)}$ for
the influenza B strains, as was found experimentally (compare the
\textit{Predicted} columns in
\fref[figTetheredAntibodyMeasurements]\letter{A-C}). For example, the two H3N2
strains (A/Panama/2007/99 and A/Wisconsin/67/05) were again found to have
measured $\text{IC}_{50}$s ($15\,\text{nM}$ and $23\,\text{nM}$) far smaller
than their predicted lower bound of $300\,\text{nM}$ (red squares in
\fref[figTetheredAntibodyModel]\letter{D}, red lines in
\fref[figTetheredAntibodyMeasurements]\letter{C}).

A second construct containing all four antibody domains attached two copies of
$\text{Ab}_{\text{A1}} \text{\textendash} \text{Ab}_{\text{A2}}
\text{\textendash} \text{Ab}_{\text{B}}^{(1)} \text{\textendash}
\text{Ab}_{\text{B}}^{(2)}$ through an IgG backbone
(\fref[figTetheredAntibodyMeasurements]\letter{D}). Since the identical domains
in both arms of this construct should be able to simultaneously bind, the new
antibody should markedly boost potency through avidity. Surprisingly, the
neutralization of this final construct was well characterized as half the
$\text{IC}_{50}$ of an individual $\text{Ab}_{\text{A1}} \text{\textendash}
\text{Ab}_{\text{A2}} \text{\textendash} \text{Ab}_{\text{B}}^{(1)}
\text{\textendash} \text{Ab}_{\text{B}}^{(2)}$, suggesting that there was no
noticeable avidity and that the increase in neutralization only arose from
having twice as many antibody domains. As above, this intriguing result presents
an opportunity to both quantitatively check experimental results and to advocate
for future studies in potentially highly promising directions. In this
particular instance, it suggests that the IgG backbone used did not permit
simultaneous binding of both arms. If a different multivalent scaffold (perhaps
with greater flexibility or with longer linkers) enabled bivalent binding of
both arms, it could potentially increase the neutralization of this construct by
100-fold as seen in the influenza B constructs.

\section*{Discussion}

In this work, we developed a statistical mechanical model that predicts the
collective efficacy of an antibody mixture whose constituents are assumed to
bind to a single site on a receptor. Each antibody is first individually
characterized by its ability to bind the receptor (through its dissociation
constant $K_D$) and inhibit activity (via its potency $\alpha$) as per
\eref[eqActivity1Ab]. Importantly, this implies that the activity of each
monoclonal antibody must be measured at a minimum of two concentrations in order
to infer both parameters, and additional measurements would further refine these
parameter values and the corresponding model predictions.

After each antibody is individually characterized, the activity of a combination
of antibodies will depend upon whether they bind independently to distinct
epitopes or compete for overlapping epitopes. Theoretical models often assume
for simplicity that all antibodies bind independently, and in the contexts where
this constraint can be experimentally imposed such models can accurately predict
the effectiveness of antibody mixtures \cite{Kong2015}. Yet when the antibody
epitopes are unknown or when a large number of antibodies are combined, it is
likely that some subset of antibodies will compete with each other while others
will bind independently, which will give rise to a markedly different response.
Our model generalized these previous results to account for antibody mixtures
where arbitrary subsets can bind independently or competitively
(\eref[eqActivity2AbDistinct][eqActivity2AbOverlapping], S1 Text Section A).

We showed that in the context of the EGFR receptor, where every pairwise
interaction was measured using surface plasmon resonance, our model is better
able to predict the efficacy of all 2-Ab and 3-Ab mixtures than a model that
assumes all antibodies bind independently or competitively
(\fref[figIndependentModel]). This suggest that mixtures of antibodies do not
exhibit large synergistic effects. More generally, similar models in the
contexts of anti-cancer drug cocktails and anti-HIV antibody mixtures also found
that the majority of cases that were described as synergistic could instead be
characterized by an independent binding model \cite{Palmer2017, Kong2015}. This
raises the possibility that synergy is more the exception then the norm, and
hence that simple models can computationally explore the full design space of
antibody combinations.

While it is often straightforward to measure the efficacy of $n$ individual
antibodies, it is more challenging to quantify all $\frac{n(n+1)}{2}$ pairwise
interactions and determine which antibodies bind independently and which compete
for an overlapping epitope. We demonstrated that after each antibody is
individually characterized, our model can be applied in reverse by using the
activity of 2-Ab mixtures to classify whether antibodies compete or bind
independently (\fref[figPredictingDistinctVersusOverlappingEpitopes]).
Surprisingly, while the resulting categorizations were much sparser than the
direct SPR measurements, the classifications produced by this method predicted
the efficacy of antibody combinations with an $R^2 = 0.90$, comparable to the
predictions made using the complete SPR results
(\fref[figIndependentModel]\letter{B}). This suggests that key features of how
antibodies interact on a molecular level can be indirectly inferred from simple
activity measurements of antibody combinations.


Modern bioengineering has opened up a new avenue of mixing antibodies by
genetically fusing different components to construct multi-domain antibodies
\cite{Spiess2015}. Such antibodies can harness multivalent interactions to
greatly increase binding avidity by over 100x (e.g. comparing the
$\text{IC}_{50}$s of the A/Wisconsin/67/05 and B/Harbin/7/94 strains of the
4-fused domains on an IgG backbone in
\fref[figTetheredAntibodyMeasurements]\letter{D} to the corresponding $\text{IC}_{50}$s for
the individual antibody domains in Panels A and B). For such constructs, the
composition of the linker can heavily influence the ability to multivalently
bind and neutralize a virus \cite{Klein2014, Einav2019}, although Laursen
\textit{et al.}~surprisingly found little variation when they modified the
length of their amino acid linker (see Table S11 in Ref~\cite{Laursen2018}). Another
curious feature of their system was that placing their linear 4-domain antibody
(\fref[figTetheredAntibodyMeasurements]\letter{C}) on an IgG backbone
(\fref[figTetheredAntibodyMeasurements]\letter{D}) only resulted in a 2x
decrease in $\text{IC}_{50}$, suggesting that the two ``arms'' of the IgG could
not simultaneously bind. We would expect that a different backbone that allows
both arms to simultaneously bind would markedly increase the neutralization
potency of this construct. In this way, quantitatively modeling these
multidomain antibodies can guide experimental efforts to design more potent
constructs.

To close, we mention that two possible avenues of future work. First, although
our model classifies antibody epitopes as either distinct or overlapping, SPR
measurements indicate that there is a continuum of possible interactions. It
would be fascinating to translate this more nuanced level of interaction into
more precise dissociation constants when two antibodies are bound. Second, while
our model focused on mixtures of antibodies, it can be applied equally well to
small molecule drugs where the number of distinct combinations may be
prohibitively large to measure experimentally but straightforward to explore
computationally.

\section*{Methods}

The coefficient of determination used to quantify how well the theoretical
predictions matched the experimental measurements
(\fref[figIndependentModel]\letter{B-D},
\fref[figPredictingDistinctVersusOverlappingEpitopes]\letter{B},
\fref[figTetheredAntibodyModel]\letter{D}) was calculated using
\begin{equation}
R^2 = 1 - \frac{\sum_{j=1}^{n} \left( y_{\text{measured}}^{(j)} - y_{\text{predicted}}^{(j)} \right)^2}{\sum_{j=1}^{n} \left( y_{\text{data}}^{(j)} \right)^2}
\end{equation}
where $y_{\text{measured}}$ and $y_{\text{predicted}}$ represent a vector of the
measured and predicted activities for the $n$ mixtures analyzed. In
\fref[figTetheredAntibodyModel]\letter{D}, we computed the $R^2$ of
$\log_{10}(\text{activity})$ to prevent the largest activities from
dominating the result (since the $\text{IC}_{50}$ values span multiple decades).

Data from the EGFR antibody mixtures was obtained by digitizing
Ref~\cite{Koefoed2011} Fig S1 using WebPlotDigitizer \cite{Rohatgi2017}. Data
for the influenza multidomain antibodies was obtained from the authors of
Ref~\cite{Laursen2018}. 

The EGFR antibody epitopes experimentally characterized through SPR
(\fref[figPredictingDistinctVersusOverlappingEpitopes]\letter{A}, bottom-left)
were categorized as overlapping if the average of the two antibody measurements
(with preincubation by either antibody) were $>50$ and as distinct if the
average was $<50$.

The original nomenclature for the antibodies used in Koefoed
\textit{et al.}~and Laursen \textit{et al.}~are given in S1 Text Table
S1.

%


%
%

%


\pagebreak
\appendix

\renewcommand{\thepage}{S\arabic{page}}
\renewcommand{\thefigure}{S\arabic{figure}}
\renewcommand{\thetable}{S\arabic{table}}
\renewcommand{\theequation}{S\arabic{equation}}
\setcounter{page}{1}
\setcounter{figure}{0}
\setcounter{table}{0}
\setcounter{equation}{0}

\section{Characterizing Antibodies Targeting the Receptor Tyrosine Kinase EGFR} \label{appendixEGFR}

\subsection{Monoclonal Antibody Binding to a Receptor} \label{appendixDerivationLigandReceptor}

In this section, we explain the states and weights notation used to develop
the equilibrium statistical mechanical models used in this work. As a focus, we
consider the case of an antibody binding to a single site on a receptor as shown
in \fref[figStatesWeightsOneLigandOneReceptor]\letter{A}. We assume that the
concentration of the antibody far exceeds that of the receptor, so that one
binding event will not noticeably affect the concentration $c$ of free antibody.

The receptor can exist in two states where it is either unbound or bound to the
antibody, where the relative probability of the bound state compared to the
unbound state equals $\frac{c}{K_D}$, where $K_D$ is the dissociation constant
of the receptor-antibody binding \cite{Bintu2005}. The (normalized) probability
of each state is given by its relative weight divided by the sum of the relative
weights of all states. For example, the probability that the receptor is bound
to the antibody is shown to be the standard $\frac{\frac{c}{K_D}}{1 +
	\frac{c}{K_D}}$ sigmoidal response. As expected, the receptor will always be
unbound in the absence of antibody ($c = 0$), while the receptor will always be
bound at saturating concentrations of antibody ($c \to \infty$).

Each receptor state also has a relative activity (i.e. the activity of the
receptor when it is in this state). In the context of EGFR, where the fractional
activity is measured relative to the receptor in the absence of antibody, the
relative activity of the unbound state is by definition equal to 1. As in the
main text, we define the activity of the bound receptor to be $\alpha$, where
$\alpha=0$ implies that the receptor is completely inactive when the antibody is bound
and $\alpha=1$ represents the opposite limit where antibody binding does not inhibit
the receptor's activity. A value of $0 < \alpha < 1$ represents an antibody that
partially inhibits activity upon binding.

The activity of the receptor in each state is given by the product of its
relative activity and the normalized probability of that state. Lastly, the
average activity of the receptor is given by the sum of its activity in each
state,
\begin{equation} \label{eqSIligandReceptorBinding}
\text{Activity} = \frac{1 + \alpha \frac{c}{K_D}}{1 + \frac{c}{K_D}}.
\end{equation}

\begin{figure}[h!]
	\centering \includegraphics{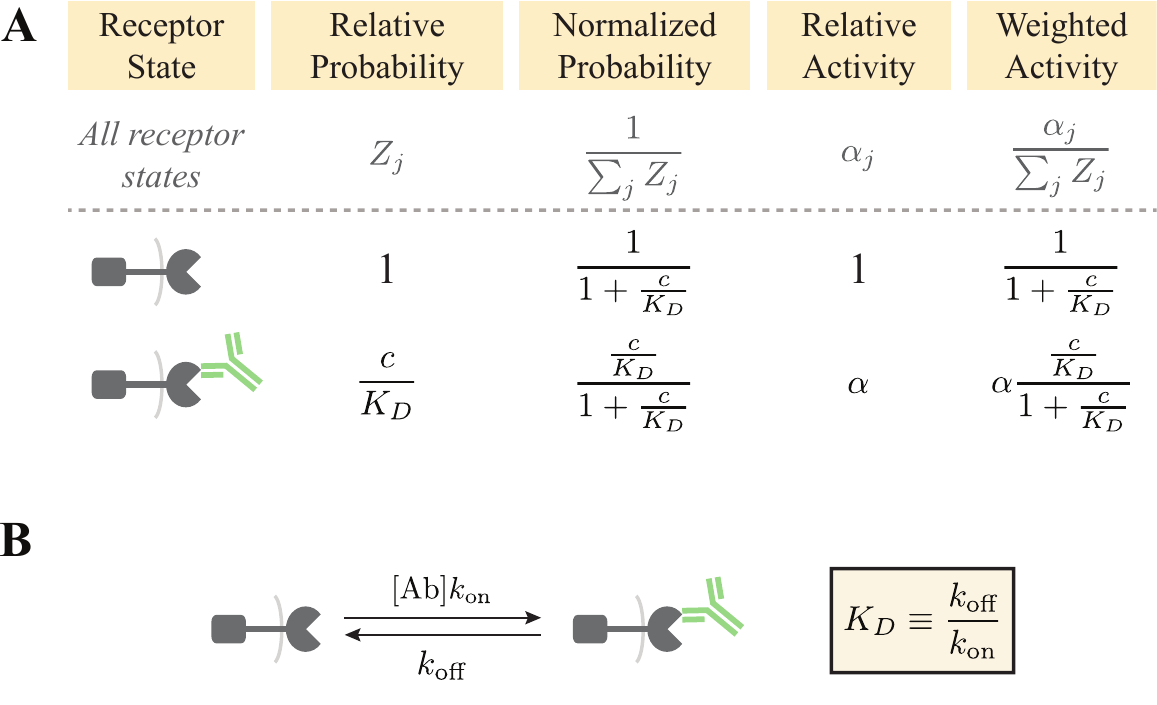}
	
	\caption{\textbf{A statistical mechanical model of an antibody binding to a
			receptor.} \letterParen{A} The relative probability and relative activity of
		each receptor state enable us to derive the average activity of the receptor.
		\letterParen{B} This method gives the same result of the dynamics model shown
		by the multiple rates, provided that the system is in steady state.}
	\label{figStatesWeightsOneLigandOneReceptor}
\end{figure}

As a quick aside, we note out that all the models considered in this work
analyze binding reactions that quickly reach steady, and hence the equilibrium
model derived above will should accurately describe such systems.
\eref[eqSIligandReceptorBinding] can also be derived by considering the dynamics
of the system shown in the rates diagram
\fref[figStatesWeightsOneLigandOneReceptor]\letter{B}. The unbound receptor
($\text{R}$) will switch to the bound state ($\text{R} \text{\textendash}
\text{Ab}$) at the rate $c k_\text{on}$ (recall that $c$ is the concentration of antibody) and subsequently unbind at a
rate $k_\text{off}$. Hence, the system is governed by the two differential
equations
\begin{align}
\frac{d[\text{R}]}{dt} &= k_\text{off} [\text{R}
\text{\textendash} \text{Ab}] - c k_\text{on} [\text{R}] \\
\frac{d[\text{R}
	\text{\textendash} \text{Ab}]}{dt} &= c k_\text{on} [\text{R}] - k_\text{off} [\text{R}
\text{\textendash} \text{Ab}].
\end{align}

Since the total amount of receptor $[\text{R}_\text{tot}] = [\text{R}] +
[\text{R} \text{\textendash} \text{Ab}]$ is fixed, these two equations are
equivalent, and either one can be solved to yield the dynamics of the system,
namely,
\begin{equation}
[R] = c_1 e^{- \left( k_\text{off} + c k_\text{on} \right) t} + [\text{R}_\text{tot}] \frac{k_\text{off}}{k_\text{off} + c k_\text{on}}.
\end{equation}
$c_1$ is fixed by the initial concentration of free receptor $[R]$ at $t=0$, but
regardless of its value, we see that the exponential term will shrink to zero at
a time scale of $\frac{1}{k_\text{off} + c k_\text{on}}$, after which the system
will be in steady state with $[R] = [\text{R}_\text{tot}]
\frac{k_\text{off}}{k_\text{off} + c k_\text{on}}$. Defining the dissociation
constant $K_D \equiv \frac{k_\text{off}}{k_\text{on}}$, the probability that any 
receptor will be unbound in steady state is given by
\begin{equation}
\frac{[R]}{[\text{R}_\text{tot}]} = \frac{1}{1 + \frac{c}{K_D}}
\end{equation}
as found in \fref[figStatesWeightsOneLigandOneReceptor]\letter{A}.

To close, we note that antibodies typically have a $K_D = 10^{-12}
\text{\textendash} 10^{-8} \, \text{M}$, and assuming a diffusion-limited on
rate $k_\text{on} = 10^{8} \text{\textendash} 10^{-9} \, \frac{1}{\text{M} \cdot
	\text{s}}$, this implies that $k_\text{off} = 10^{-3} \text{\textendash} 10^{1}
\, \frac{1}{\text{s}}$. Therefore, an upper bound (in the limit of little
antibody) for the time scale it takes for such a system to reach steady state is
$10^{-1} \text{\textendash} 10^{3} \, \text{s}$, which is the time before
experimental measurements should be conducted. Since antibodies are typically
preincubated for even longer periods during experiments, an equilibrium model
should be valid in all of the case studies we consider in this work.

\subsection{The Fractional Activity of 3-Ab Mixtures} \label{appendix3AbMixtures}

As described in the main text, the fractional activity of 2-Ab mixtures is given
by Eq 2 if the two antibodies bind to distinct epitopes
and Eq 3 if they bind to overlapping epitopes. These
equations are straightforward to extend to mixtures with multiple antibodies by
drawing all of the statistical weights and Boltzmann weights for the mixture
(analogous to Fig 1) and then computing the fractional
activity as per Section \ref{appendixDerivationLigandReceptor}.

For example, a mixture of three antibodies all binding to distinct epitopes
would give rise to
\begin{equation} \label{eqActivity3AbDistinct}
\text{Fractional Activity}_{\text{(1,2,3 distinct epitopes)}} = \left( \frac{1 + \alpha_1 \frac{c_1}{K_D^{(1)}}}{1 + \frac{c_1}{K_D^{(1)}}} \right) \left( \frac{1 + \alpha_2 \frac{c_2}{K_D^{(2)}}}{1 + \frac{c_2}{K_D^{(2)}}} \right) \left( \frac{1 + \alpha_3 \frac{c_3}{K_D^{(3)}}}{1 + \frac{c_3}{K_D^{(3)}}} \right).
\end{equation}
If antibodies 1 and 2 bind to an overlapping epitope but antibody 3 binds to a
distinct epitope, then
\begin{equation} \label{eqActivity3AbDistinctAndCompetitive}
\text{Fractional Activity}_{\text{(1,2 overlapping epitopes; 3 distinct epitope)}} = \left( \frac{1 + \alpha_1 \frac{c_1}{K_D^{(1)}} + \alpha_2 \frac{c_2}{K_D^{(2)}}}{1 + \frac{c_1}{K_D^{(1)}} + \frac{c_2}{K_D^{(2)}}} \right) \left( \frac{1 + \alpha_3 \frac{c_3}{K_D^{(3)}}}{1 + \frac{c_3}{K_D^{(3)}}} \right).
\end{equation}
If all three antibodies binds to overlapping epitopes, the fractional activity becomes
\begin{equation} \label{eqActivity3AbCompetitive}
\text{Fractional Activity}_{\text{(1,2 overlapping epitopes; 3 distinct epitope)}} = \frac{1 + \alpha_1 \frac{c_1}{K_D^{(1)}} + \alpha_2 \frac{c_2}{K_D^{(2)}} + \alpha_3 \frac{c_3}{K_D^{(3)}}}{1 + \frac{c_1}{K_D^{(1)}} + \frac{c_2}{K_D^{(2)}} + \frac{c_3}{K_D^{(3)}}}.
\end{equation}

\subsection{Characterizing Ten EGFR Monoclonal Antibodies from Koefoed 2011} \label{appendixmAbCharacterizations}

Koefoed \textit{et al.}~investigated how a panel of ten monoclonal antibodies
inhibit EGFR by measuring the protein's activity at multiple antibody
concentrations in the human cell line HN5 \cite{Koefoed2011}. By fitting these
titration curves to Eq 1, we can infer the dissociation constant
$K_D$ between each antibody and EGFR as well as the potency $\alpha$ of each
antibody in the HN5 cell line (\fref[figCharacterizingKoefoed2011]\letter{A,B}).
For each curve, the $K_D$ corresponds to the midpoint of the curve (halfway
between its minimum and maximum activity values) while $\alpha$ represents the
activity at saturating antibody concentration.

Koefoed \textit{et al.}~found that the majority (7/10) of these
antibodies reduced EGFR activity below 20\% at saturating concentrations, and
since mixtures of antibodies would likely further decrease activity their
potency would be difficult to accurately measure. To that end, Koefoed
\textit{et al.}~switched to the A431NS cell line that is partially resistant to
EGFR antibodiess where they remeasured EGFR activity for all ten monoclonal antibodies as well as
mixtures of two or three Abs (with 1:1 and 1:1:1 ratios, respectively).
Each measurement was performed at a mixture concentration of $2 \frac{\mu
	\text{g}}{\text{mL}}$, implying that each antibody was half as dilute in the
2-Ab mixtures and one-third as dilute in the 3-Ab mixtures relative to the monoclonal antibody
measurement.

\begin{figure}[b!]
	\centering \includegraphics{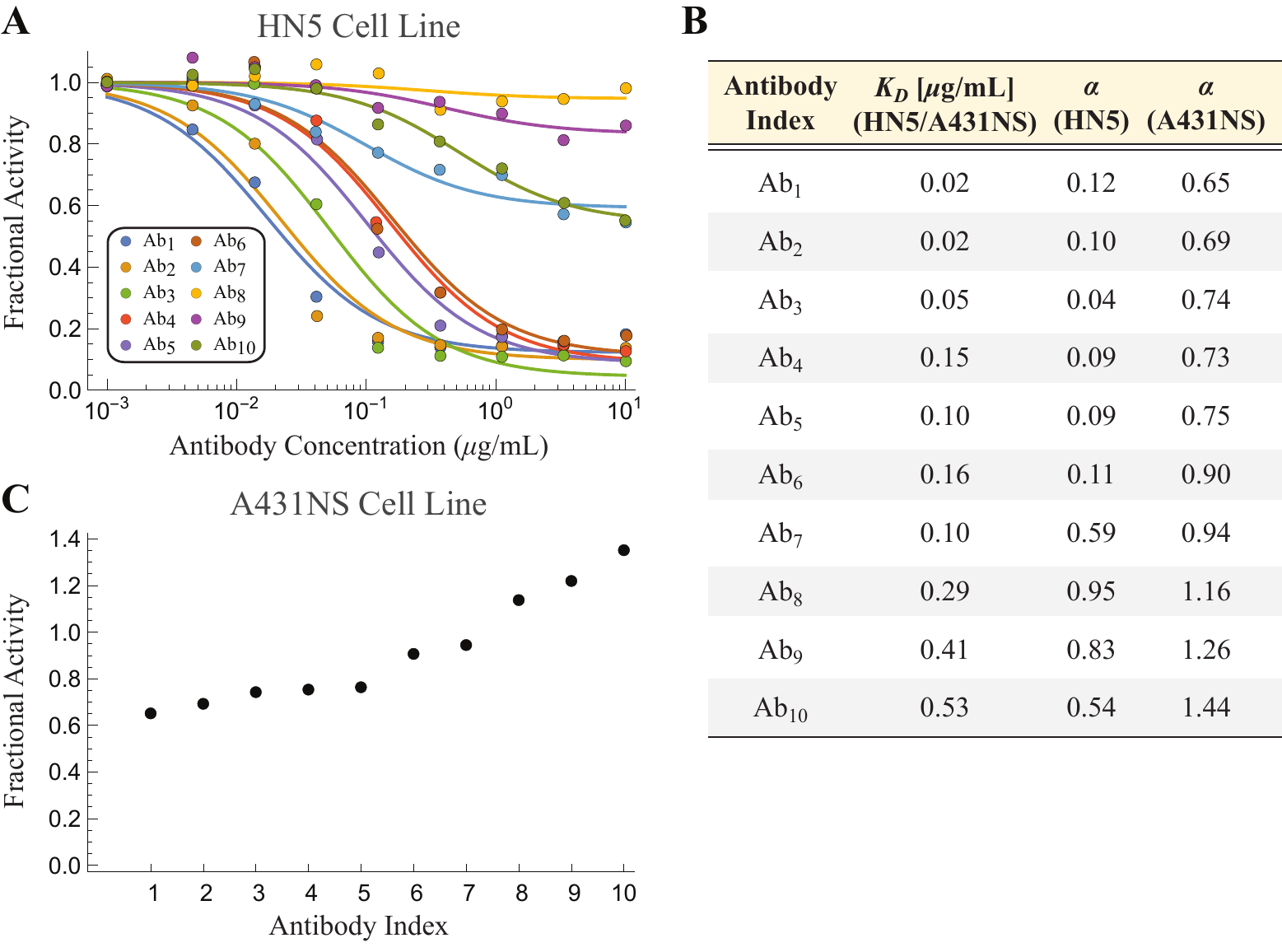}
	
	\caption{\textbf{Inferring the model parameters for the 10 monoclonal antibodies in Koefoed
			2011.} \letterParen{A} Activity of EGFR in the presence of ten antibodies in the HN5
		cell line. \letterParen{B} The inferred dissociation constant ($K_D$) and
		fractional activity in the presence of saturating antibody ($\alpha$) in each
		cell line. \letterParen{C} Fractional EGFR activity in the presence of $2
		\frac{\mu \text{g}}{\text{mL}}$ of each antibody in the A431NS cell line. Data
		reproduced from Ref~\cite{Koefoed2011} Figure 2B,C.}
	\label{figCharacterizingKoefoed2011}
\end{figure}

We assume that the antibody-EGFR binding interaction is identical in the A431NS cell
line, and hence that these same $K_D$ parameters characterizes these antibodies in
that cell line. Hence, by using the additional measurement of each antibody's
potency at $2 \frac{\mu \text{g}}{\text{mL}}$ in the A431NS cell line
(\fref[figCharacterizingKoefoed2011]\letter{C}), we can infer the potency
$\alpha$ of each antibody in the A431NS cell line
(\fref[figCharacterizingKoefoed2011]\letter{B}). These parameters are sufficient
to predict how any mixture (at any concentration and ratio) will behave
in the HN5 and A431NS cell lines. Note that all data presented in the main text
correspond to the A431NS cell line.

\subsection{Comparing the HN5 and A431NS Cell Lines from Koefoed 2011} \label{appendixKoefoedCellLines}

Koefoed \textit{et al.}~measured the potency of 176 mixtures in the A431NS cell
line but only 55 mixtures in the HN5 cell line.
\fref[figKoefoed2011BothCellLines] extends our analysis to both the A431NS and
HN5 cell lines. While the majority of mixtures have very little predicted and
measured activity ($\lesssim 0.2$), approximately 13 outliers fall outside this
range and appear to be poorly predicted.

While the coefficient of determination $R^2 = 0.61$ is significantly lower for
this cell line, we note that: (1) there are far fewer data points in this cell
line and (2) that our $R^2$ definition places more importance on points with
larger predicted or measured fractional activity, and hence these few outliers
have a disproportionate effect. That said, it remains unknown whether with more
data our model would be as successful in the HN5 cell line. Another open
question is why some antibody mixtures had inhibitory effects in one cell line
but exacerbating effects in the other (e.g. the mixture of
$\text{Ab}_3+\text{Ab}_{10}$ resulted in 1.22 fractional activity of EGFR in the
A431NS cell line but 0.19 fractional activity in the HN5 cell line).

\begin{figure}[h!]
	\centering \includegraphics{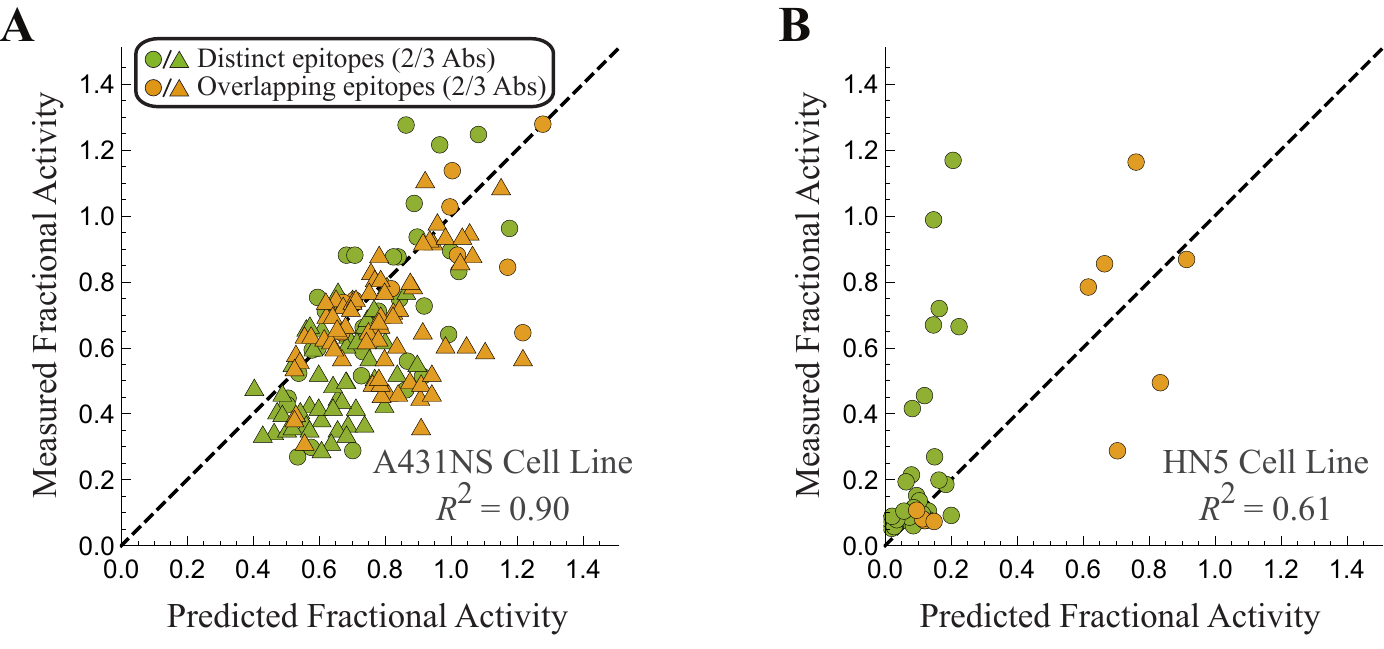}
	
	\caption{\textbf{Predicting the potency of antibody mixtures on EGFR in
			different cell lines.} Our model predictions versus experimental measurements
		for EGFR activity in the \letterParen{A} A431NS and \letterParen{B} HN5 cell
		lines.} \label{figKoefoed2011BothCellLines}
\end{figure}

\subsection{Separating the 2-Ab and 3-Ab Predictions for EGFR Antibody Mixtures} \label{appendixEGFRseparatePredictions}

\fref[figEGFRseperatePredictions] separates the 2-Ab and 3-Ab mixture
predictions from the three models in Fig 2. More
specifically, Panels \letter{A} and \letter{B} of
\fref[figEGFRseperatePredictions] show the predictions for combinations of two
and three antibodies using the epitope mappings produced by SPR (see captions on
the diagonal of Fig 2\letter{A}; there are four EGFR
epitopes bound by antibodies \#1-3, \#4, \#5-6, and \#7-10, respectively).

Without this SPR data, we could have alternately assumed that antibodies all
bind independently (\fref[figEGFRseperatePredictions]\letter{C,D}) or that all
antibodies vie for the same epitope
(\fref[figEGFRseperatePredictions]\letter{E,F}). Either of these models generate
slightly worse predictions, as exhibited by their lower coefficients of
determination $R^2$.

\begin{figure}[b!]
	\centering \includegraphics{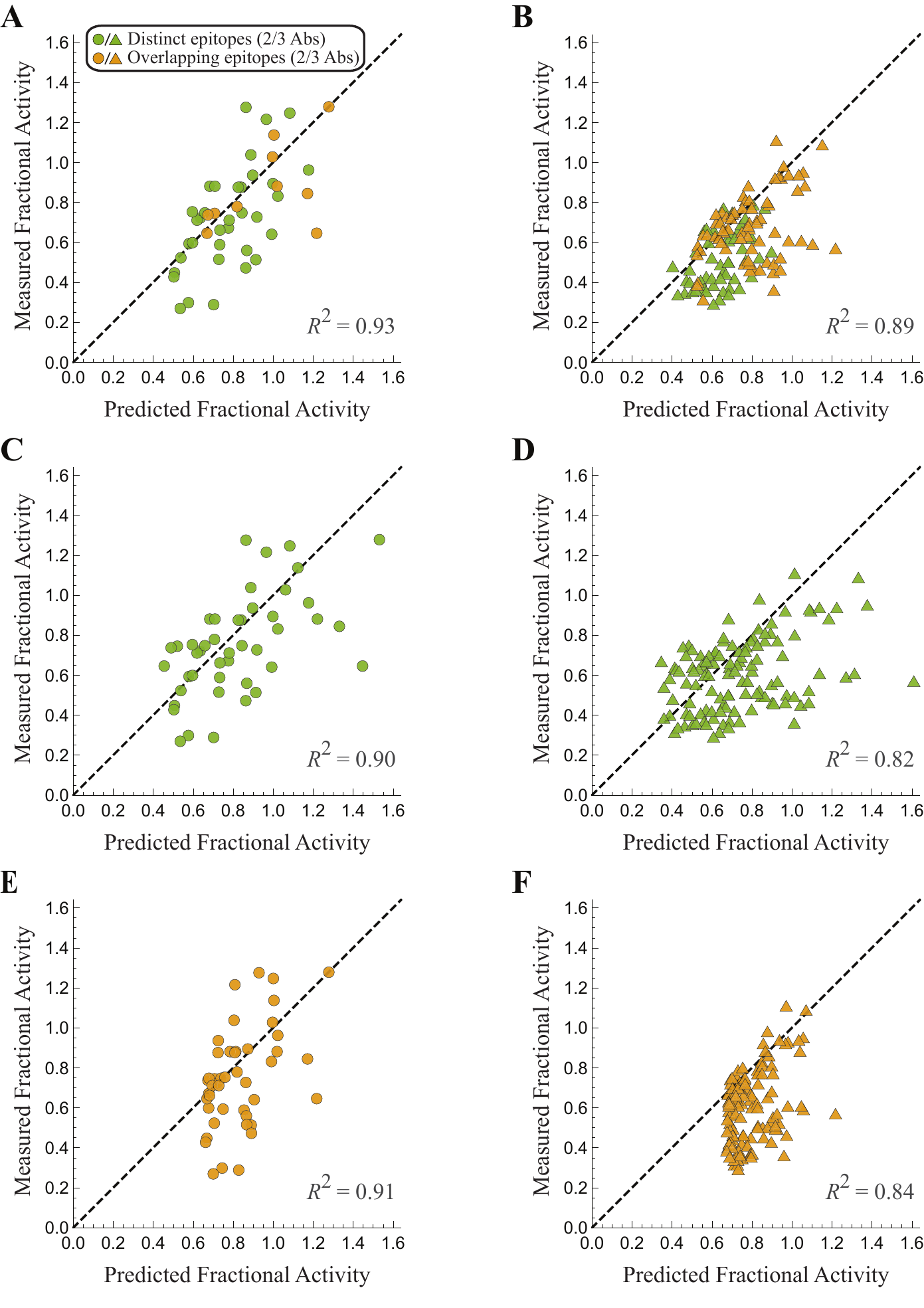}
	
	\caption{\textbf{Separate predictions for 2-Ab and 3-Ab mixtures.}
		\letterParen{A,B} Predictions using the epitope mappings produced using SPR.
		\letterParen{C,D} Predictions assuming that all antibodies bind independently.
		\letterParen{E,F} Predictions assuming that all antibodies bind competitively.}
	\label{figEGFRseperatePredictions}
\end{figure}

%
%
%
%

\pagebreak
\subsection{Original Antibody Nomenclature}

Table \ref{tablemAbNomenclature} shows the antibodies considered in this work
were named in the original works of Refs.~\cite{Koefoed2011} and
\cite{Laursen2018}. In our work, we indexed the EGFR antibodies by their potency in
the A431NS cell line and labeled the influenza antibodies by the viral group
that they most effectively neutralized.

\begin{table}[h!]
	\begin{center}
		\begin{tabular}{ccccc}
			\hhline{--~--}
			\multicolumn{2}{c}{\cellcolor[HTML]{f2e8d4}\textbf{Koefoed 2011 Antibody Nomenclature}} & & \multicolumn{2}{c}{\cellcolor[HTML]{f2e8d4}\textbf{Laursen 2018 Antibody Nomenclature}} \\ 
			\hhline{==~==}
			\cellcolor[HTML]{fff4df}\textit{\textcolor[HTML]{fff4df}{.....}This Work\textcolor[HTML]{fff4df}{.....}} & \cellcolor[HTML]{fff4df}\textit{Original Work}	&	& \cellcolor[HTML]{fff4df}\textit{\textcolor[HTML]{fff4df}{.....}This Work\textcolor[HTML]{fff4df}{.....}} & \cellcolor[HTML]{fff4df}\textit{Original Work}   \\
			\hhline{--~--}
			$\text{Ab}_1$                               & 1565	&	& $\text{Ab}_{\text{A1}}$ & SD38		\\
			\cellcolor[HTML]{f0f0f0}$\text{Ab}_2$       & \cellcolor[HTML]{f0f0f0}1320 & & \cellcolor[HTML]{f0f0f0}$\text{Ab}_{\text{A2}}$ & \cellcolor[HTML]{f0f0f0}SD36		\\
			$\text{Ab}_3$                               & 1024	&	& $\text{Ab}_{\text{B}}^{(1)}$ & SD83	\\
			\cellcolor[HTML]{f0f0f0}$\text{Ab}_4$       & \cellcolor[HTML]{f0f0f0}992  & & \cellcolor[HTML]{f0f0f0}$\text{Ab}_{\text{B}}^{(2)}$ & \cellcolor[HTML]{f0f0f0}SD84	\\
			\hhline{~~~--}
			$\text{Ab}_5$                               & 1277								\\
			\cellcolor[HTML]{f0f0f0}$\text{Ab}_6$       & \cellcolor[HTML]{f0f0f0}1030		\\
			$\text{Ab}_7$                               & 1284								\\
			\cellcolor[HTML]{f0f0f0}$\text{Ab}_8$       & \cellcolor[HTML]{f0f0f0}1347		\\
			$\text{Ab}_9$                               & 1260								\\
			\cellcolor[HTML]{f0f0f0}$\text{Ab}_{10}$    & \cellcolor[HTML]{f0f0f0}1261		\\
			\hhline{--}\\
		\end{tabular}
		\caption{\textbf{Matching the antibody nomenclature used in this work with the names in the original manuscript.}}
		\label{tablemAbNomenclature}
	\end{center}
\end{table}

\pagebreak
\section{Characterizing Distinct versus Overlapping EGFR Epitopes} \label{appendixCharacterizingEpitopes}

In this section, we describe in more detail how we can use activity data from
the 2-Ab mixtures to classify which subsets of antibodies bind to the same
epitopes. A basic classification scheme would categorize two antibodies as binding to
distinct epitopes if Eq 2 predicted their mixture's
activity better than Eq 3; otherwise, the two antibodies
would be categorized as binding to overlapping epitopes.

However, this simple classification scheme does not account for the uncertainty
that arises from experimental noise. As an extreme case, if the activities of
a mixtures are measured at extremely small concentrations, then the
fractional activity will be $\approx 1$ for all mixtures (as will be the
predictions from both the distinct and overlapping models), and this
classification scheme would only be fitting the noise. Hence, it is best to
measure the activity of each mixture at saturating antibody concentrations where
the signal-to-noise of the system will be greatest.

We determined from Koefoed \textit{et al.}~(Figure S1) that the standard error
of the mean (SEM) of their activity measurements was $\sigma = 0.04$, and we
proceed to incorporate this uncertainty into our categorization scheme using a
simple threshold model. More specifically, we add two components to the
classification scheme: (1) As shown in
\fref[figAntibodyMixtureBayesModified]\letter{A}, activity measurements that
fall within $\sigma$ of the midpoint of the two model predictions are left
unclassified, since experimental noise could easily lead to such points being
incorrectly classified. (2) If the two model predictions lie sufficiently close
(within $4\sigma$) to one another, then the uncertainty (from both the
measurement and the model predictions) make it difficult to distinguish between
the two models with certainty. 

As mentioned above, for potent antibodies that are measured at saturating
concentrations, the difference between the independent and overlapping binding
models will be large, making it easier to definitively classify antibody
epitopes. Koefoed \textit{et al.}~measured their 2-Ab combinations at a total
concentration of $2\,\frac{\mu \text{g}}{\text{mL}}$ (and hence $1\,\frac{\mu
	\text{g}}{\text{mL}}$ for each antibody), and
\fref[figCharacterizingKoefoed2011]\letter{A} suggest that while this is close
to saturating concentration for the majority of antibodies, increasing the
concentration by 10x may have allowed more pairs of antibodies to be
definitively categorized.

\begin{figure}[b!]
	\centering \includegraphics{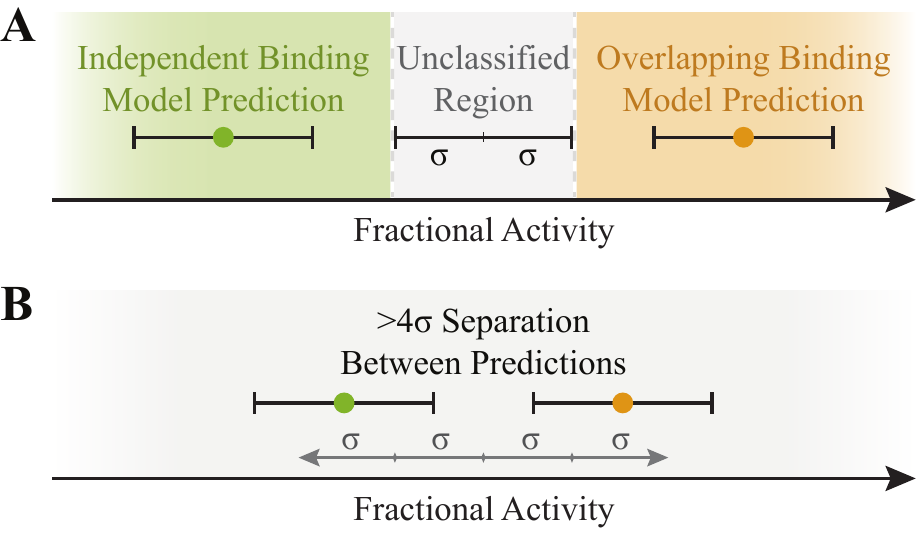}
	
	\caption{\textbf{Accounting for uncertainty in the classification scheme.}
		\letterParen{A} 2-Ab combinations whose activity falls within $\sigma$ of the
		midpoint between the two model predictions will be labeled as ``unclassified,''
		since the difference in distance from the measurement to either prediction is
		less than the experimental error. \letterParen{B} In addition, when the two
		model predictions are less than $4\sigma$ different, any measurement is labeled
		as ``unclassified'' because measurement error could result in overlap with both
		model predictions.} \label{figAntibodyMixtureBayesModified}
\end{figure}

\section{Characterizing Multidomain Antibodies} \label{appendixLaursen2018TetheredAntibodies}

\subsection{Relating Influenza Neutralization to Binding}

In this section, we discuss how the microscopic dissociation constants of two
tethered antibodies (shown in Fig 4\letter{C}) relate
to influenza viral neutralization. We begin by considering a single antibody
with an effective dissociation constant $K_D^{(1)}$ that quantifies its avidity
to the virus \cite{Xu2016}. The probability that this antibody at concentration
$c$ will be bound to a virion is given by
\begin{equation}
p_\text{bound} = \frac{\frac{c}{K_D^{(1)}}}{1 + \frac{c}{K_D^{(1)}}}.
\end{equation}
For influenza virus with $N \approx 400$ hemagglutinin (HA) trimers, the number
of bound trimers is given by $N_\text{bound} = N p_\text{bound}$.

The relationship between viral binding and neutralization remain unclear
\cite{Klasse2002}. It has been suggested that $\sim 50$ HA trimers are required
to infect a cell \cite{Xu2016}. However, an IgG bound to one trimer may
sterically preclude neighboring HA from binding. It has been proposed that
neutralization is a sigmoidal function of neutralization (see Figure S1 of
Ref~\cite{Ndifon2009}),
\begin{equation} \label{eqFractionNeutralizationGeneral}
\text{Fraction Neutralized} = \frac{N^h + N_{50}^h}{N^h} \frac{\left( N p_\text{bound} \right)^h}{\left( N p_\text{bound} \right)^h + N_{50}^h},
\end{equation}
where $N_{50}$ is the number of bound trimers required to reduce infectivity to
50\%, $h$ is a Hill coefficient, and the prefactor assures that the fraction
neutralized ranges from 0 (in the absence of antibody) to 1 (in the presence of
saturating antibody).

In the absence of data for our influenza strain of interest, we will assume
$h=1$ in the following analysis. This enables us to rewrite
\eref[eqFractionNeutralizationGeneral] as
\begin{equation}
\text{Fraction Neutralized} =  \frac{\frac{c}{\text{IC}_{50}^{(1)}}}{1 + \frac{c}{\text{IC}_{50}^{(1)}}},
\end{equation}
where we have defined the inhibitory concentration of antibody at which 50\% of
the virus is neutralized
\begin{equation} \label{eqKDandIC50relation}
\text{IC}_{50}^{(1)} = \frac{N_{50}}{N + N_{50}} K_D^{(1)}.
\end{equation}
For example, if the virus is 50\% neutralized when $N_{50} = 100$ trimers are
bound, the midpoint of a viral neutralization curve would occur at roughly
\nicefrac{1}{5} the antibody concentration required to bind 50\% of the trimers,
as has been observed for some influenza antibodies (see Figure 2 of
Ref~\cite{Knossow2002}).

We now consider the tethered two-domain antibody shown in Fig 4\letter{C}.
Denote the antibody concentration as $c$, the effective dissociation constants
of its two domains as $K_D^{(1)}$ and $K_D^{(2)}$, and the effective
concentration when both domains simultaneously bind as $\tilde{c}_{\text{eff}}$
(which we will shortly relate to the $c_{\text{eff}}$ in the antibody
neutralization given in Eq 4). The probability that this antibody is bound to a
virion is given by
\begin{equation} \label{eqPBoundMultidomainAntibody}
p_\text{bound} = \frac{\frac{c}{K_D^{(1)}} + \frac{c}{K_D^{(2)}} + \frac{c}{K_D^{(1)}} \frac{\tilde{c}_{\text{eff}}}{K_D^{(1)}}}{1 + \frac{c}{K_D^{(1)}}}.
\end{equation}

As above, we assume that neutralization is related to the binding probability
through \eref[eqFractionNeutralizationGeneral] with Hill coefficient $h=1$,
which upon substituting \eref[eqPBoundMultidomainAntibody] yields
\begin{equation} \label{eqAppendixNeutralization}
\text{Fraction Neutralized} = \frac{\frac{c}{\text{IC}_{50,\text{A1}}} + \frac{c}{\text{IC}_{50,\text{A2}}} +
	\frac{c}{\text{IC}_{50,\text{A1}}} \frac{c_{\text{eff}}}{\text{IC}_{50,\text{A2}}}}{1 +
	\frac{c}{\text{IC}_{50,\text{A1}}} + \frac{c}{\text{IC}_{50,\text{A2}}} +
	\frac{c}{\text{IC}_{50,\text{A1}}} \frac{c_{\text{eff}}}{\text{IC}_{50,\text{A2}}}}
\end{equation}
where we have defined the $\text{IC}_{50}$s of both antibodies using
\eref[eqKDandIC50relation] as well as the rescaled effective concentration
\begin{equation}
c_{\text{eff}} = \frac{N_{50}}{N + N_{50}} \tilde{c}_{\text{eff}}.
\end{equation}
Therefore, in the case where $h=1$ the functional form of neutralization
(\eref[eqAppendixNeutralization]) is identical to the probability that an HA
trimer is bound (\eref[eqPBoundMultidomainAntibody]), with the dissociation
constants and the effective concentration rescaled by $\frac{N_{50}}{N +
	N_{50}}$.

To put these results into perspective, we note that many viruses are covered in
spikes (analogous to influenza HA) that enable them to bind and fuse to their
target cells \cite{Klasse2002}, and hence the sigmoidal dependence between viral
binding and neutralization is likely widely applicable. However, HIV is a clear
exception, since each virion has an average of 14 envelope spikes
\cite{Klein2010}. In that context, neutralization is roughly proportional to the
number of bound spikes so that $\text{IC}_{50} \approx K_D$
\cite{Brandenberg2017}.

\subsection{Assuming Different Antibody Constructs have Distinct Effective Concentrations} \label{appendixDifferentEffectiveConcentrations}

In the main text, we quantified the boost in avidity from tethering two antibodies using
the effective concentration $c_{\text{eff}} = 1400\,\text{nM}$ by using
least-squares regression to minimize the (log) predicted $\text{IC}_{50}$ for
each tethered construct binding to all strains. This effective concentration
depends on the distance between binding sites on a virion, and hence the tethered
$\text{Ab}_{\text{A1}} \text{\textendash} \text{Ab}_{\text{A2}}$ construct may have
a different $c_{\text{eff}}$ when binding to influenza A group 1 and group 2
strains, and $\text{Ab}_{\text{B}}^{(1)} \text{\textendash}
\text{Ab}_{\text{B}}^{(2)}$ may have yet another effective concentration
when binding to the influenza B strains.

\fref[figFusedAntibodiesIndividualCEff] shows the best-fit $c_{\text{eff}}$ for
each of these cases. While this plot suggests that there are differences between
each tethered antibody and influenza strain, we note that there is very limited
data to infer such values (e.g., there are 7, 7, and 5 data points in the
influenza A1, A2, and B groups, respectively; note that we ignore the H3N2
outlier strains A/Panama/2007/99 and A/Wisconsin/67/05 discussed in the main
text). That said, incorporating this fine-grained level of modeling could
further boost the accuracy of modeling efforts and is worth pursuing as more
data is gathered.

Finally, we mention that Laursen \textit{et al.} measured the efficacy of
$\text{Ab}_{\text{A1}} \text{\textendash} \text{Ab}_{\text{A2}}$ with different
linkers of length 18 amino acids ($\sim 63 \text{\AA}$), 38 amino acids ($\sim
133 \text{\AA}$), and 60 amino acids ($\sim 210 \text{\AA}$) against the four
influenza strains H1N1 A/California/07/09, H1N1 A/Puerto Rico/8/34-MA, H5N1
A/Vietnam/1194/04, and H3N2 A/Wisconsin/67/05 (see Ref~\cite{Laursen2018} Table
S11). They found very little difference between the $\text{IC}_{50}$ of each
construct, which might naively suggest that the length of the linker does not
matter. However, since these four strains were all negligibly inhibited by
$\text{Ab}_{\text{A2}}$ ($\text{IC}_{50} \ge 1000\,\text{nM}$; see
Fig 5\letter{A}), this domain cannot
meaningfully contribute to bivalent binding, so that $\text{Ab}_{\text{A1}}
\text{\textendash} \text{Ab}_{\text{A2}}$ would be expected to be as potent as
$\text{Ab}_{\text{A1}}$ irrespective of linker length. On the other hand, the
length of the linker should matter when both Abs in a multidomain antibody can
bind, as is the case for $\text{Ab}_{\text{A1}} \text{\textendash}
\text{Ab}_{\text{A2}}$ binding to the three influenza A group 2 strains with
$\text{IC}_{50} < 1000\,\text{nM}$ and for $\text{Ab}_{\text{B}}^{(1)}
\text{\textendash} \text{Ab}_{\text{B}}^{(2)}$ binding to the five influenza B
strains.

\begin{figure}[h]
	\centering \includegraphics{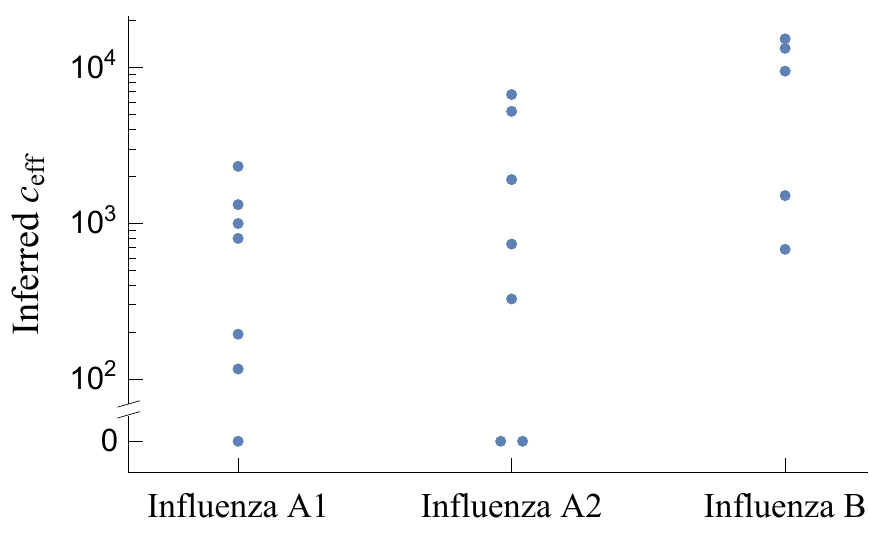}
	
	\caption{\textbf{Inferring $c_{\text{eff}}$ between every tethered construct
			binding to each group of influenza virus.} The best-fit effective concentration
		of $\text{Ab}_{\text{A1}} \text{\textendash} \text{Ab}_{\text{A2}}$
		against the influenza A group 1 and group 2 strains, together with the
		effective concentration of $\text{Ab}_{\text{B}}^{(1)} \text{\textendash}
		\text{Ab}_{\text{B}}^{(2)}$ binding to influenza B.}
	\label{figFusedAntibodiesIndividualCEff}
\end{figure}

\pagebreak


\begin{thebibliography}{10}
	
	\bibitem{Awwad2018}
	Awwad S, Angkawinitwong U.
	\newblock {Overview of Antibody Drug Delivery}.
	\newblock Pharmaceutics. 2018;10(3):83.
	\newblock doi:{10.3390/pharmaceutics10030083}.
	
	\bibitem{Caskey2019}
	Caskey M, Klein F, Nussenzweig MC.
	\newblock {Broadly neutralizing anti-HIV-1 monoclonal antibodies in the
		clinic}.
	\newblock Nature Medicine. 2019;25(4):547--553.
	\newblock doi:{10.1038/s41591-019-0412-8}.
	
	\bibitem{Perelson1997}
	Perelson AS, Weisbuch G.
	\newblock {Immunology for physicists}.
	\newblock Reviews of Modern Physics. 1997;69(4):1219--1268.
	\newblock doi:{10.1103/RevModPhys.69.1219}.
	
	\bibitem{Mukherjee2010}
	Mukherjee S.
	\newblock {The emperor of all maladies: A biography of cancer}.
	\newblock Scribner; 2010.
	
	\bibitem{Chow2013}
	Chow SK, Smith C, MacCarthy T, Pohl MA, Bergman A, Casadevall A.
	\newblock {Disease-enhancing antibodies improve the efficacy of bacterial
		toxin-neutralizing antibodies}.
	\newblock Cell Host {\&} Microbe. 2013;13(4):417--28.
	\newblock doi:{10.1016/j.chom.2013.03.001}.
	
	\bibitem{Koefoed2011}
	Koefoed K, Steinaa L, Soderberg JN, Kjaer I, Jacobsen HJ, Meijer PJ, et~al.
	\newblock {Rational identification of an optimal antibody mixture for targeting
		the epidermal growth factor receptor}.
	\newblock mAbs. 2011;3(6):584--595.
	\newblock doi:{10.4161/mabs.3.6.17955}.
	
	\bibitem{Laursen2018}
	Laursen NS, Friesen RHE, Zhu X, Jongeneelen M, Blokland S, Vermond J, et~al.
	\newblock {Universal protection against influenza infection by a multidomain
		antibody to influenza hemagglutinin}.
	\newblock Science. 2018;362(6414):598--602.
	\newblock doi:{10.1126/science.aaq0620}.
	
	\bibitem{Spiess2015}
	Spiess C, Zhai Q.
	\newblock {Alternative molecular formats and therapeutic applications for
		bispecific antibodies}.
	\newblock Molecular Immunology. 2015;67(2):95--106.
	\newblock doi:{10.1016/J.MOLIMM.2015.01.003}.
	
	\bibitem{Knossow2002}
	Knossow M, Gaudier M, Douglas A, Barr{\`{e}}re B, Bizebard T, Barbey C, et~al.
	\newblock {Mechanism of Neutralization of Influenza Virus Infectivity by
		Antibodies}.
	\newblock Virology. 2002;302(2):294--298.
	\newblock doi:{10.1006/VIRO.2002.1625}.
	
	\bibitem{Ndifon2009}
	Ndifon W, Wingreen NS, Levin SA.
	\newblock {Differential Neutralization Efficiency of Hemagglutinin Epitopes,
		Antibody Interference, and the Design of Influenza Vaccines}.
	\newblock Proceedings of the National Academy of Sciences of the United States
	of America. 2009;106(21):8701--6.
	\newblock doi:{10.1073/pnas.0903427106}.
	
	\bibitem{Einav2019}
	Einav T, Yazdi S, Coey A, Bjorkman PJ, Phillips R.
	\newblock {Harnessing Avidity: Quantifying Entropic and Energetic Effects of
		Linker Length and Rigidity Required for Multivalent Binding of Antibodies to
		HIV-1 Spikes}.
	\newblock bioRxiv. 2019; p. 406454.
	\newblock doi:{10.1101/406454}.
	
	\bibitem{Kong2015}
	Kong R, Louder MK, Wagh K, Bailer RT, DeCamp A, Greene K, et~al.
	\newblock {Improving neutralization potency and breadth by combining broadly
		reactive HIV-1 antibodies targeting major neutralization epitopes.}
	\newblock Journal of virology. 2015;89(5):2659--71.
	\newblock doi:{10.1128/JVI.03136-14}.
	
	\bibitem{Palmer2017}
	Palmer AC, Sorger PK.
	\newblock {Combination Cancer Therapy Can Confer Benefit via Patient-to-Patient
		Variability without Drug Additivity or Synergy}.
	\newblock Cell. 2017;171(7):1678--1691.e13.
	\newblock doi:{10.1016/J.CELL.2017.11.009}.
	
	\bibitem{Klein2014}
	Klein JS, Jiang S, Galimidi RP, Keeffe JR, Bjorkman PJ, Regan L.
	\newblock {Design and characterization of structured protein linkers with
		differing flexibilities}.
	\newblock Protein Engineering, Design and Selection. 2014;27(10):325--330.
	\newblock doi:{10.1093/protein/gzu043}.
	
	\bibitem{Rohatgi2017}
	Rohatgi A. {WebPlotDigitizer}; 2017.
	\newblock Available from: \url{https://automeris.io/WebPlotDigitizer}.
	
	\bibitem{Bintu2005}
	Bintu L, Buchler NE, Garcia HG, Gerland U, Hwa T, Kondev J, et~al.
	\newblock {Transcriptional regulation by the numbers: models}.
	\newblock Current Opinion in Genetics {\&} Development. 2005;15(2):116--124.
	\newblock doi:{10.1016/j.gde.2005.02.007}.
	
	\bibitem{Xu2016}
	Xu H, Shaw DE.
	\newblock {A Simple Model of Multivalent Adhesion and Its Application to
		Influenza Infection}.
	\newblock Biophysical Journal. 2016;110(1):218--233.
	\newblock doi:{10.1016/j.bpj.2015.10.045}.
	
	\bibitem{Klasse2002}
	Klasse PJ, Sattentau QJ.
	\newblock {Occupancy and Mechanism in Antibody-Mediated Neutralization of
		Animal Viruses}.
	\newblock Journal of General Virology. 2002;83(9):2091--2108.
	\newblock doi:{10.1099/0022-1317-83-9-2091}.
	
	\bibitem{Klein2010}
	Klein JS, Bjorkman PJ.
	\newblock {Few and Far Between: How HIV May Be Evading Antibody Avidity}.
	\newblock PLoS Pathogens. 2010;6(5):e1000908.
	\newblock doi:{10.1371/journal.ppat.1000908}.
	
	\bibitem{Brandenberg2017}
	Brandenberg OF, Magnus C, Rusert P, G{\"{u}}nthard HF, Regoes RR, Trkola A.
	\newblock {Predicting HIV-1 transmission and antibody neutralization efficacy
		in vivo from stoichiometric parameters}.
	\newblock PLoS Pathogens. 2017;13(5):1--35.
	\newblock doi:{10.1371/journal.ppat.1006313}.
	
\end{thebibliography}
\end{document}